\documentclass[12pt,reqno]{amsart}

\usepackage{amsmath,amssymb,amsfonts,amsthm}
\usepackage{bm,bbm}
\usepackage{comment}

\usepackage{caption}
\usepackage{placeins}
\usepackage{graphicx}
\usepackage{subcaption}
\usepackage{booktabs}
\usepackage{float}
\usepackage{color}
\usepackage{placeins}
\usepackage[colorlinks=true,allcolors=blue]{hyperref}

\usepackage{enumitem}

\usepackage{algorithm}
\usepackage{algpseudocode}

\newtheorem{thm}{Theorem}[section]
\newtheorem{lem}[thm]{Lemma}
\newtheorem{remark}{Remark}

\usepackage{fullpage}

\begin{document}
\title{Reconstruction methods for inverse scattering problems with phaseless data}
\author{John C. Schotland}
\address{Department of Mathematics and Department of Physics, Yale University, New Haven, CT, USA}
\email{john.schotland@yale.edu}
\author{Shenwen Yu}
\address{Department of Mathematical Sciences, Tsinghua University, Beijing, China}
\email{ysw22@mails.tsinghua.edu.cn}
\begin{abstract}
We investigate phaseless inverse scattering problems for the Schr\"odinger equation and develop reconstruction methods based on the inverse Born series (IBS). 
We consider three types of phaseless data: the far-field total field, the total field and the far-field scattered field. For phaseless total-field data, we extend the IBS framework and analyze its radius of convergence. 
In the far-field setting, we propose a Fourier-based reconstruction method that exploits the scattering reciprocity between incident and observation directions to recover Fourier coefficients of the scattering potential. For phaseless scattered-field data, we employ a polarization-based approach to recover phase information and enable IBS reconstruction. Numerical experiments are conducted to validate the proposed methods, demonstrating their effectiveness for low-contrast scattering potentials and their limitations in high-contrast cases.
\end{abstract}
\maketitle
 
\section{Introduction} 

We consider the Schr\"odinger equation
\begin{equation*}
\label{schrodinger}
    -\Delta u + V(\boldsymbol x) u = k^2 u \quad \text{in } \mathbb{R}^d,
\end{equation*}
where the field $u$ is generally complex--valued and the potential $V\in C_c^\infty(\mathbb R^d;\mathbb R)$.
The scattered field $u_s$ is taken to obey the radiation condition
\begin{equation*}
    \lim_{r \to \infty} r^{\frac{d-1}{2}}
    \left( \frac{\partial u_s}{\partial r} - \mathrm{i}k\, u_s \right) = 0,
\end{equation*}
where $r = |\boldsymbol x|$. This paper is concerned with the inverse problem of recovering $V$ from 
phaseless measurements.  We consider several variants of this problem, with data consisting of measurements of $|u|$,  as well as far-field measurements of $|u|$ and  $|u_s|$. 

Inverse scattering problems play a central role in nearly every branch of
physics. A comprehensive theory is available for the conventional problem 
in which both the modulus and phase of the field is known~\cite{coltonInverseAcousticElectromagnetic2019, isakovInverseProblemsPartial2017}. In many practical applications, however, phase information is difficult or even impossible to acquire, and one has access only to intensity measurements. This gives
rise to phaseless inverse scattering problems. We emphasize that the phaseless setting is significantly more challenging than conventional inverse scattering problems due to the sublinear nature of the absolute--value function and related matters of non-uniqueness.

Historically, the study of phaseless inverse problems can be traced to the
one-dimensional setting, where uniqueness and reconstruction results were
first established. See, for example, Refs.~\cite{%
sacksRecoverySingularitiesAmplitude1997,
sacksInverseProblemCoupled2004,
novikovPhaselessInverseScattering2015,
klibanovPhaselessInverseScattering1992,
klibanovPhaseRetrievalProblem1995,
aktosunInverseProblemLine1998}.
In higher dimensions, several classes of phaseless problems can
be distinguished. For the case of the phaseless total field,  uniqueness results and numerical methods have been established~\cite{%
novikovHolographicUniquenessTheorem2024,
novikovFormulasPhaseRecovering2015, klibanovNumericalMethodSolve2018}.
In addition, high-frequency
approximations have been studied
in~\cite{klibanovTwoReconstructionProcedures2015}.

A second approach focuses on the phaseless scattered
field. A common strategy is to make use of superpositions of incident waves, including plane waves and
point sources. Related uniqueness results for such problems have been reported
in~\cite{%
zhangUniqueDeterminationsInverse2020,
xuUniquenessInverseAcoustic2020,
sunUniquenessPhaselessInverse2019,
romanovPhaselessInverseProblems2018}.
Related work on numerical methods includes boundary integral equation
approaches~\cite{zhangFastImagingScattering2018}, optimization~\cite{zhangRecoveringScatteringObstacles2017}.
A related line of work introduces known reference scatterers or sources, often together with superposed incident waves, to provide additional phase information. Related uniqueness results can be found in~\cite{%
zhangUniquenessInverseCavity2020,
zhangUniquenessResultsPhaseless2018},
while related numerical methods are developed in~\cite{%
dongReferenceBallBased2018,
agaltsovIterativeApproachMonochromatic2018,
zhangReconstructionAcousticSources2020,
hohagePhaseRetrievalPhaseless2024}.
Finally, we note that the optical theorem provides an alternative approach to the phaseless inverse problem~\cite{govyadinovPhaselessThreeDimensionalOptical2009,carminati_schotland_book}. This approach has been applied to inverse problems with incident fields consisting of a superposition of plane waves and to problems in which a reference scatterer is present.

In this work, we investigate reconstruction methods for two phaseless inverse scattering
problems. The first problem concerns phaseless measurements of the total field, while the
second concerns phaseless measurements of the scattered field. Both problems are studied
within the framework of the inverse Born series (IBS), which represents the solution of the
inverse problem as an explicitly computable series in the measured data. The IBS has been
extensively studied for conventional inverse problems in which phase information is available;
see~\cite{moskow12InverseBorn2019} for a review.

Compared with previous approaches to phaseless inverse problems, the present
work introduces new reconstruction strategies for the two types of measurements.
For phaseless total-field data, to the best of our knowledge, this is the first
work to apply the IBS directly to the phaseless problem by
deriving a Born series for the intensity measurements. Moreover, in the far-field setting, we develop a new
treatment of the associated linearized inverse problem by interchanging the
incident and observation directions, which allows the relevant Fourier
coefficients of the scattering potential to be recovered explicitly.

For phaseless scattered-field data, in contrast to the direct superpositions
of incident waves used in related work, we employ two incident plane waves
with coefficients $a\in\{\pm1,\pm i\}$. In particular, the use of the complex
coefficients $\pm i$ enables us to apply the polarization identity and recover
complex-valued scattering information for the subsequent IBS
reconstruction.

This paper is organized as follows. In
Section~\ref{sec:pre}, we introduce the definition and analysis of the IBS. In
Section~\ref{sec:total}, we derive the inversion formulas and analyze
the convergence of the Born series and the IBS for phaseless
total-field data. We also provide a  justification of the Fourier
method. In Section~\ref{sub:polar}, we study the phaseless scattered-field
problem and analyze the polarization method. In
Sections~\ref{sec:ag} and~\ref{sec:result}, we present the numerical
algorithms and demonstrate their performance in numerical experiments.
Finally, our conclusions are presented  in Section~\ref{sec:con}.

\section{Preliminaries}\label{sec:pre}
\subsection{Far-field Approximation}\label{sub:farfield}
The total field is written as
\begin{equation*}
    u(\boldsymbol x) = u_0(\boldsymbol x) + u_s(\boldsymbol x),
\end{equation*}
where $u_0$ is the incident field and $u_s$ is the scattered field.
In the absence of the scatterer, $u_0$ satisfies the free-space
Schrödinger equation. The scattered field obeys 
\begin{equation*}
    -\Delta u_s + V(\boldsymbol x) u = k^2 u_s .
\end{equation*}
It follows that $u$ obeys the Lippmann--Schwinger equation
\begin{equation*}
    u(\boldsymbol x) = u_0(\boldsymbol x) + \int_{\mathbb{R}^d}
    G(\boldsymbol x, \boldsymbol y) V(\boldsymbol y) u(\boldsymbol y)\,d\boldsymbol y ,
\end{equation*}
where  $G(\boldsymbol x, \boldsymbol y)$ is the free-space Green's function~\cite{coltonInverseAcousticElectromagnetic2019}.

The Green's function is given by
\[
\begin{aligned}
G(\boldsymbol x, \boldsymbol y) &=\frac{\mathrm{i}}{4} H_0^{(1)}(k|\boldsymbol x- \boldsymbol y|), \quad d=2, \\
G(\boldsymbol x, \boldsymbol y) &=\frac{e^{\mathrm{i}k|\boldsymbol x- \boldsymbol y|}}{4\pi|\boldsymbol x- \boldsymbol y|}, \quad d=3 .
\end{aligned}
\]
When $|\boldsymbol x|\gg |y|$, the Green's function admits the asymptotic expansion
\begin{equation*}
    G(\boldsymbol x, \boldsymbol y) \sim C_d
    \frac{e^{\mathrm{i}k|\boldsymbol x|}}{|\boldsymbol x|^{\frac{d-1}{2}}}
    e^{-\mathrm{i}k\,\hat{\boldsymbol x}\cdot \boldsymbol y},
\end{equation*}
where $\hat{\boldsymbol x}=x/|\boldsymbol x|$ and
\begin{equation*}
   C_d=2^{-\frac{d+1}{2}} \pi^{-\frac{d-1}{2}} k^{\frac{d-3}{2}}
   \exp\!\left(-\mathrm{i}\frac{\pi}{4}(d-3)\right).
\end{equation*}
Consequently, the scattered field has the far-field asymptotic form
\begin{equation*}
    u_s(\boldsymbol x) \sim C_d 
    \frac{e^{\mathrm{i}k|\boldsymbol x|}}{|\boldsymbol x|^{\frac{d-1}{2}}}
    A(\hat{\boldsymbol x}),
\end{equation*}
where the scattering amplitude is given by
\begin{equation}\label{farfield}
    A(\hat{\boldsymbol x}) =
    \int_{\mathbb{R}^d}
    e^{-\mathrm{i}k\,\hat{\boldsymbol x}\cdot \boldsymbol y}
    V(\boldsymbol y) u(\boldsymbol y)\,d\boldsymbol y.
\end{equation}

In the Born approximation, we replace the total field $u$ in \eqref{farfield} by the incident field $u_0(\boldsymbol y)=e^{\mathrm{i}\boldsymbol{k}'\cdot \boldsymbol y}$. This yields
\begin{equation*}
    A_B(\boldsymbol k,\boldsymbol k') =
    \int_{\mathbb{R}^d}
    e^{-\mathrm{i}(\boldsymbol k-\boldsymbol k')\cdot \boldsymbol y}
    V(\boldsymbol y)\,d\boldsymbol y,
\end{equation*}
where
\[
\boldsymbol k = k\hat{\boldsymbol x}, \quad |\boldsymbol k| = |\boldsymbol k'| = k .
\]
Therefore,
\[
A_B(\boldsymbol k,\boldsymbol k') = \widehat{V}(\boldsymbol k-\boldsymbol k'),
\]
where $\widehat{V}$ denotes the Fourier transform of the potential $V$. In particular, when $V$ is nonnegative, we note that
\begin{equation*}
    \widehat{V}(0) = \int_{\mathbb{R}^d} V(\boldsymbol x)\,d\boldsymbol x \ge 0 .
\end{equation*}

\subsection{Inverse Born Series}
We employ IBS in the form described in \cite{hoskinsAnalysisInverseBorn2022}.
Let $X$ and $Y$ be Banach spaces and let $K_m : X^m \rightarrow Y$ be $m$-multilinear operators for $m \ge 1$. Consider the operator $\mathcal{F}:X\rightarrow Y$ defined by the forward series
\begin{equation}\label{forward formulation}
    \mathcal{F}[\eta] = \sum_{m=1}^{\infty} K_m(\eta,\ldots,\eta).
\end{equation}

The $K_m$ are referred to as forward operators. The forward problem consists of evaluating the map $\mathcal{F}: \eta \mapsto \phi$ for $\eta \in X$ and $\phi \in Y$.

The inverse problem is to recover $\eta$ from $\phi$. To this end, we seek an operator $\mathcal{I}:Y\rightarrow X$ which acts as an inverse of $\mathcal{F}$ in an appropriate sense. Formally, $\mathcal{I}$ is defined by the inverse series
\begin{equation}\label{inverse formulation}
    \mathcal{I}[\phi] = \sum_{m=1}^{\infty} \mathcal{K}_m(\phi).
\end{equation}

By substituting \eqref{inverse formulation} into \eqref{forward formulation},
it can be shown in \cite{hoskinsAnalysisInverseBorn2022} that the inverse operators $\{\mathcal{K}_m\}_{m\ge1}$ are defined recursively by
\begin{align*}
\mathcal{K}_1(\phi) &= K_1^{+}(\phi), \\
\mathcal{K}_2(\phi) &= -\mathcal{K}_1\!\left(
K_2\!\left(\mathcal{K}_1(\phi),\mathcal{K}_1(\phi)\right)
\right), \\
\mathcal{K}_m(\phi) &=
-\sum_{n=2}^{m}
\sum_{i_1+\cdots+i_n=m}
\mathcal{K}_1
K_n\!\left(
\mathcal{K}_{i_1}(\phi),\ldots,\mathcal{K}_{i_n}(\phi)
\right).
\end{align*}

Here $K_1^{+}$ denotes the bounded inverse of $K_1$ when it exists, and otherwise its pseudoinverse.

We next recall the convergence radius and an error estimate for the IBS as described in \cite{hoskinsAnalysisInverseBorn2022}.

\begin{thm}\label{thm:convergent}
Let $\mu$ and $\nu$ be positive constants. Suppose that
\[
\|K_m(\eta_1,\ldots,\eta_m)\|_Y
\le
\nu \mu^{m-1}
\|\eta_1\|_X \cdots \|\eta_m\|_X,
\quad m=1,2,\ldots .
\]

Then the IBS converges if $\|\mathcal{K}_1 \phi\|_X < r$, where the radius of convergence $r$ is given by
\[
r=\frac{1}{2\mu}
\left[\sqrt{16C^2+1}-4C\right],
\]
with
\[
C=\max\{2,\|\mathcal{K}_1\|\nu\}.
\]

Moreover, if $\mathcal{K}_1\phi \in B_r$, then the inverse operator $\mathcal{I}$ maps $B_r$ into $B_{r_0}$ with
\[
r_0=\frac{2\mu}{\sqrt{16C^2+1}}.
\]
\end{thm}

\begin{thm}
Suppose that the hypotheses of Theorem~\ref{thm:convergent} hold and that both the forward Born series and the IBS converge. Let $\tilde{\eta}$ denote the sum of the IBS and define $\eta_1=\mathcal{K}_1\phi$. Set
\[
\mathcal{M}:=\max\{\|\eta\|_X,\|\tilde{\eta}\|_X\},
\]
and assume that
\[
\mathcal{M}
<
\frac{1}{\mu}
\left(
1-\sqrt{\frac{\nu\|\mathcal{K}_1\|}{1+\nu\|\mathcal{K}_1\|}}
\right).
\]

Then the approximation error satisfies
\[
\begin{aligned}
\left\|
\eta-\sum_{m=1}^N \mathcal{K}_m(\phi)
\right\|_X
\le
&\,M
\left(\frac{\|\eta_1\|_X}{r}\right)^{N+1}
\frac{1}{1-\frac{\|\eta_1\|_X}{r}} \\
&+
\left(
1-\frac{\nu\|\mathcal{K}_1\|}{(1-\mu\mathcal{M})^2}
+\nu\|\mathcal{K}_1\|
\right)^{-1}
\|
(I-\mathcal{K}_1K_1)\eta
\|_X ,
\end{aligned}
\]
where
\[
M=\frac{2\mu}{\sqrt{16C^2+1}}.
\]
\end{thm}
\begin{remark}
    Theorem~2.2 makes explicit the role of the linearized inversion in the
    accuracy of the IBS reconstruction. The reconstruction error depends on
    the stability of the inverse, or pseudoinverse, of $K_1$. Consequently,
    better conditioning of the linearized inverse problem leads to better
    control of the error in the inverse Born series reconstruction.
\end{remark}
\section{Phaseless Total Field}\label{sec:total}

In this section we consider the inverse problem when phaseless measurements of the total field are available. We first derive the Born series and the corresponding IBS in this setting. 
This framework applies to both general and far-field measurements. 
When only far-field data are available, we further introduce a Fourier-based method to efficiently solve the linearized problem. This step plays a crucial role in the practical implementation of the IBS. 

\subsection{Born Series}

Suppose that only the magnitude of the total field $|u|$ is measured. 
We introduce the Born series expansion
\begin{equation}\label{def of original born}
    u = u_0 + u_1 + u_2 + \cdots ,
\end{equation}
where
\begin{equation*}
u_m(\boldsymbol x)
=
\int_{\mathbb{R}^d} G(\boldsymbol x,\boldsymbol  y_1)V(\boldsymbol y_1)
\int_{\mathbb{R}^d} G(\boldsymbol y_1,\boldsymbol y_2)V(\boldsymbol y_2)
\cdots
\int_{\mathbb{R}^d} G(\boldsymbol y_{m-1},\boldsymbol y_m)V(\boldsymbol y_m)u_0(\boldsymbol y_m)
\,d\boldsymbol y_1\cdots d\boldsymbol y_m .
\end{equation*}
The Born series expansion is obtained by iterating the Lippmann--Schwinger equation. This yields:
\begin{equation}\label{def of phase born}
    u
    =
    K^p_1(V)
    +
    K^p_2(V,V)
    + \cdots ,
\end{equation}
where 
\begin{equation}\label{Forward of phase}
    K^p_m(V_1,\dots,V_m)=\int_{\mathbb{R}^d} G(\boldsymbol x,\boldsymbol y_1)V_1(\boldsymbol y_1)\cdots
\int_{\mathbb{R}^d} G(\boldsymbol y_{m-1},\boldsymbol y_m)V_m(\boldsymbol y_m)u_0(\boldsymbol y_m)
\,d\boldsymbol y_1\cdots d\boldsymbol y_m .
\end{equation}
To obtain the Born series for phaseless data, we expand the squared magnitude of the field, which yields
\begin{equation}\label{def of born}
    |u|^2 - |u_0|^2
    =
    K_1(V)
    +
    K_2(V,V)
    + \cdots ,
\end{equation}
where
\begin{equation}\label{def of Km}
    K_m(V,\dots,V)
    =
    \sum_{j=0}^{m}
    u_j\,u_{m-j}^* .
\end{equation}

It will prove useful to introduce a sequence of  potentials $(V_1,\dots,V_n)$. 
For $i\ge1$, define
\[
u_i(V_1,\dots,V_i)(\boldsymbol x)
=
\int_{\Omega} G(\boldsymbol x,\boldsymbol y_1)V_1(\boldsymbol y_1)
\cdots
\int_{\Omega} G(\boldsymbol y_{i-1}, \boldsymbol y_i)V_i( \boldsymbol y_i)u_0( \boldsymbol y_i)
\,d\boldsymbol y_i\cdots d\boldsymbol y_1,
\]
and set $u_0(V_1,\dots,V_0)\equiv u_0$. Here we assume that $V$ is compactly supported in $\Omega$. 
Then the $n$th-order multilinear term in the forward series is
\[
K_n(V_1,\dots,V_n)(\boldsymbol x_d)
=
\sum_{i=0}^{n}
u_i(V_1,\dots,V_i)(\boldsymbol x_d)\,
u^*_{n-i}(V_{i+1},\dots,V_n)(\boldsymbol x_d),
\]
where by convention $u_0(V_{n+1},\dots,V_n)\equiv u_0$.

Suppose that there are $l$ incident waves and the detectors are placed on $\omega$, where $\omega=\partial\Omega$ for the
near-field setting and $\omega$ is a circle sufficiently far from $\Omega$ for the far-field setting.
For any multilinear operator $K$ of order $n$ on $\left(L^{\infty}(\Omega)\right)^{n}$, we define
\[
|K|_{\infty}
=
\sup_{V_1,\ldots,V_n \neq 0}
\frac{
\left\|K\left(V_1,\ldots,V_n\right)\right\|_{\left(C(\omega)\right)^l}
}{
\left\|V_1\right\|_{\infty}\cdots\left\|V_n\right\|_{\infty}
}.
\]
\begin{lem}
Define
\[
\mu_0
=
\sup_{x\in\Omega\cup\omega}
\int_{\Omega} |G(\boldsymbol x, \boldsymbol y)|\,d\boldsymbol y .
\]
Then
\[
|K_m|_{\infty}
\le
(m+1)\mu_0^{m}\|u_0\|_{L^\infty}^2 .
\]
Furthermore, assume that $d=2$ and the spatial domain $\Omega$ is contained in a disk with radius $R>1$. Then we have the estimate:

$$
\mu_0\leq\frac{k^2}{4}\left(I(k)
    +\frac{4}{3} \sqrt{\frac{2 \pi}{k}}\big(R^{3 / 2}-1\big)\right)
$$

where 
\[
I(k) :=
\begin{cases}
\pi + 2\left(\gamma+\log\frac{k}{2}-\tfrac{1}{2}\right) + \dfrac{8}{k^2}e^{-2\gamma}, & k \;\geq\; 2e^{-\gamma}, \\[1.2em]
(\pi+1) - 2\gamma - 2\log\frac{k}{2}, & k \;<\; 2e^{-\gamma},
\end{cases}
\]
and $\gamma$ is the Euler constant.
\end{lem}

\begin{thm}
Let $\mu = 2\mu_0$ and $\nu = \|u_0\|_{L^\infty(\Omega)}^2$. 
Then,
\[
|K_m|_{\infty}
\le
\mu^m \nu,
\quad m\in\mathbb{N}^* .
\]
\end{thm}

The proofs of the above results follow directly from the definitions of $u_m$ and $K_m$, together with repeated applications of the triangle inequality, and are therefore omitted.

\begin{remark}
For inverse scattering with phase information, the corresponding bounds for the Born series are $\mu=\mu_0$ and $\nu=\|u_0\|_{L^\infty(\Omega)}$. Detailed calculations for $\mu_0$ in 2 and 3  dimensional space can be found in \cite{hoskinsAnalysisInverseBorn2022, schotlandInverseScatteringDirac2026}.
\end{remark}

Since the operator $K_1$ is explicitly defined in \eqref{def of Km}, its pseudoinverse $\mathcal{K}_1$ can be implemented numerically, following the approach in \cite{defilippisBornInverseBorn2023,defilippisNonlinearityHelpsConvergence2024}.

\subsection{Fourier Method}\label{sub:fourier method}

In this subsection, the measured quantity remains the phaseless total field
considered in the previous subsection. We now assume, however, that the
measurements are taken sufficiently far from the scattering region so that
the far-field approximation can be applied to the scattered component of the
total field. This leads to a Fourier-based approach for constructing
$\mathcal{K}_1$. The key idea is to recover the Fourier coefficients of the
potential $V$ by exchanging the incident and observation directions.

It follows from \eqref{def of born} that the linear term in the forward series is given by
\[
u_0u_1^*+u_1u_0^*.
\]
For convenience, we denote this linearized intensity term by
\[
D(\boldsymbol x,\boldsymbol k')=u_0(\boldsymbol x)u_1(\boldsymbol x)^*+u_1(\boldsymbol x)u_0(\boldsymbol x)^*,
\qquad
u_0(\boldsymbol x)=e^{i\boldsymbol k'\cdot \boldsymbol x}.
\]
Applying the far-field approximation (1) to $u_1$ and taking
$k'=k\hat y$, we obtain
\begin{align*}
D(\boldsymbol x,\boldsymbol k')
&=
e^{\mathrm{i}\boldsymbol k'\cdot \boldsymbol x}
C^*_d\frac{e^{-\mathrm{i}k|\boldsymbol x|}}{|\boldsymbol x|^{\frac{d-1}{2}}}\widehat{V}^*(\boldsymbol k-\boldsymbol k')
+
e^{-\mathrm{i}\boldsymbol k'\cdot \boldsymbol x}
C_d
\frac{e^{\mathrm{i}k|\boldsymbol x|}}{|\boldsymbol x|^{\frac{d-1}{2}}}\widehat{V}(\boldsymbol k-\boldsymbol k') \\
&=
e^{\mathrm{i}\boldsymbol k'\cdot \boldsymbol x}
C_d^*
\frac{e^{-\mathrm{i}k|\boldsymbol x|}}{|\boldsymbol x|^{\frac{d-1}{2}}}
\widehat{V}(\boldsymbol k'-\boldsymbol k)
+
e^{-\mathrm{i}\boldsymbol k'\cdot \boldsymbol x}
C_d
\frac{e^{\mathrm{i}k|\boldsymbol x|}}{|\boldsymbol x|^{\frac{d-1}{2}}}
\widehat{V}(\boldsymbol k-\boldsymbol k').
\end{align*}

Writing $\boldsymbol x=r_x\hat{\boldsymbol x}$ gives
\begin{align*}
D(\boldsymbol x,\boldsymbol k')
&=
C_d^*
\frac{e^{\mathrm{i}kr_x(\hat{\boldsymbol x}\cdot\hat{\boldsymbol y}-1)}}{r_x^{\frac{d-1}{2}}}
\widehat{V}(k(\hat{\boldsymbol y}-\hat{\boldsymbol x}))
+
C_d
\frac{e^{\mathrm{i}kr_x(1-\hat{\boldsymbol x}\cdot\hat{\boldsymbol y})}}{r_x^{\frac{d-1}{2}}}
\widehat{V}(k(\hat{\boldsymbol x}-\hat{\boldsymbol y})).
\end{align*}

Thus
\begin{equation}\label{component 1}
D(\boldsymbol x,k\hat{\boldsymbol y})\,r_x^{\frac{d-1}{2}}
=
C_d^* e^{\mathrm{i}kr_x(\hat{\boldsymbol x}\cdot\hat{\boldsymbol y}-1)}\widehat{V}(k(\hat{\boldsymbol y}-\hat{\boldsymbol x}))
+
C_d e^{\mathrm{i}kr_x(1-\hat{\boldsymbol x}\cdot\hat{\boldsymbol y})}\widehat{V}(k(\hat{\boldsymbol x}-\hat{\boldsymbol y})).
\end{equation}

Now choose another measurement point $\boldsymbol y\in\partial\Omega$, with $\boldsymbol y=r_y\hat {\boldsymbol y}$, and take the incident wave to be
\[
u_0(\boldsymbol z)=e^{\mathrm{i}k\hat{\boldsymbol x}\cdot \boldsymbol z}.
\]
That is, we interchange the direction of incidence and the observation direction, and obtain
\begin{equation}\label{component 2}
D(\boldsymbol y,k\hat{\boldsymbol x})\,r_y^{\frac{d-1}{2}}
=
C_d^* e^{\mathrm{i}kr_y(\hat{\boldsymbol x}\cdot\hat{\boldsymbol y}-1)}\widehat{V}(k(\hat{\boldsymbol x}-\hat{\boldsymbol y}))
+
C_d e^{\mathrm{i}kr_y(1-\hat{\boldsymbol x}\cdot\hat{\boldsymbol y})}\widehat{V}(k(\hat{\boldsymbol y}-\hat{\boldsymbol x})).
\end{equation}

Combining \eqref{component 1} and \eqref{component 2} yields a linear system that allows us to recover
$\widehat{V}(k(\hat{\boldsymbol x}-\hat{\boldsymbol y}))$ and $\widehat{V}(k(\hat{\boldsymbol y}-\hat{\boldsymbol x}))$
from the measurements $D(\boldsymbol x,k\hat{\boldsymbol y})$ and $D(\boldsymbol y,k\hat{\boldsymbol x})$, provided that
\[
(C^*_d)^2e^{\mathrm{i}k(r_x+r_y)(\hat{\boldsymbol x}\cdot\hat{\boldsymbol y}-1)}
\neq
C_d^2e^{\mathrm{i}k(r_x+r_y)(1-\hat{\boldsymbol x}\cdot\hat{\boldsymbol y})}.
\]
From the definition of $C_d$ we obtain
\[
\frac{C_d^2}{(C^*_d)^2}
=
\begin{cases}
-1, & 2\mid d, \\
1, & 2\nmid d .
\end{cases}
\]

Using this procedure, we can recover the Fourier coefficients of $V$ at a collection of frequencies, after which standard reconstruction techniques can be applied to reconstruct $V$.

We remark that the linear system arising from \eqref{component 1} and \eqref{component 2} may become severely ill-conditioned. To ensure numerical stability, the corresponding frequencies are discarded in such cases. This step is incorporated into the numerical algorithm.

\section{Phaseless scattered field}\label{sub:polar}
In this section, we consider the case where phaseless measurements of the scattered field are available. We further assume that $V$ is nonnegative.
Since $|\widehat{V}|$ alone does not uniquely determine $V$, additional data are required. To overcome this difficulty, we consider incident fields of the form
\begin{equation*}
    u_0(\boldsymbol x) = e^{-\mathrm{i}\boldsymbol k_1\cdot \boldsymbol x}
    + a e^{-\mathrm{i}\boldsymbol k_2\cdot \boldsymbol x},
\end{equation*}
where
\[
|\boldsymbol k_1| = |\boldsymbol k_2| = k,
\quad
a \in \{ \pm 1, \pm \mathrm{i} \}.
\]
It follows that the scattering amplitude in the Born approximation is given by
\begin{equation*}
    A_B(\boldsymbol k_1,\boldsymbol k_2)
    = \widehat{V}(\boldsymbol k_1-\boldsymbol k_2) + a \widehat{V}(0).
\end{equation*}
Using the polarization identity
\begin{align*}
    z_1 z_2^*
    = \frac{1}{4} \Big(
        |z_1+z_2|^2 - |z_1-z_2|^2
        + \mathrm{i}|z_1+\mathrm{i}z_2|^2
        - \mathrm{i}|z_1-\mathrm{i}z_2|^2
    \Big), \quad z_1,z_2 \in \mathbb C
\end{align*}
and setting
\[
z_1 = \widehat{V}(\boldsymbol k_1-\boldsymbol k_2),
\quad
z_2 = \widehat{V}(0) \in \mathbb{R},
\]
we obtain
\begin{align*}
    \widehat{V}(\boldsymbol k_1-\boldsymbol k_2)\,\widehat{V}(0)
    = \frac{1}{4}\Big(
    &|\widehat{V}(\boldsymbol k_1-\boldsymbol k_2)+\widehat{V}(0)|^2
    - |\widehat{V}(\boldsymbol k_1-\boldsymbol k_2)-\widehat{V}(0)|^2 \\
    &+ \mathrm{i}|\widehat{V}(\boldsymbol k_1-\boldsymbol k_2)+\mathrm{i}\widehat{V}(0)|^2
    - \mathrm{i}|\widehat{V}(\boldsymbol k_1-\boldsymbol k_2)-\mathrm{i}\widehat{V}(0)|^2
    \Big).
\end{align*}
Therefore, $\widehat{V}(\boldsymbol k_1-\boldsymbol k_2)$ can be determined by performing four experiments corresponding to the choices $a\in\{\pm1,\pm i\}$. Once $\widehat{V}$ is obtained, the potential $V$ can be reconstructed in the usual way via the inverse Fourier transform:
\begin{equation*}
    V(\boldsymbol x)
    =
    \int_{|\boldsymbol k|\le 2k}
    \frac{d\boldsymbol k}{(2\pi)^d}
    e^{-\mathrm{i}\boldsymbol k\cdot \boldsymbol x}
    \widehat{V}_B(\boldsymbol k).
\end{equation*}

\begin{remark}
We note that the polarization method is motivated by the generalized optical theorem of scattering theory, as discussed in \cite{carminati_schotland_book}, Section 11.2.
This method differs from conventional phase retrieval, whose goal is to recover the missing phase of the measured data, a problem that is severely ill-posed, as discussed in \cite{barnettGeometryPhaseRetrieval2022}. Rather than recovering the phase explicitly, we incorporate additional physical measurements, namely data generated by superpositions of incident waves. While these measurements may increase the experimental complexity of data acquisition, they markedly improve the stability of the reconstruction. 
\end{remark}

\section{Numerical Algorithms}\label{sec:ag}
In this section, we present the details of our numerical algorithms, including both the forward and inverse solvers.
\subsection{Forward Solver}
Following standard procedures \cite{bergExtendedContrastSource1999}, 
we introduce the auxiliary field
\[
\sigma(\boldsymbol x) = k^2 V(\boldsymbol x)\,u(\boldsymbol x).
\]
The scattered field can then be represented as the convolution
\begin{equation}\label{eq:us_conv_sigma}
u_s(\boldsymbol x) = \int_\Omega G(\boldsymbol x- \boldsymbol y)\,\sigma(\boldsymbol y)\,d\boldsymbol y .
\end{equation}
Since $V$ is compactly supported in $\Omega$, the density $\sigma$ has the same support.
Substituting \eqref{eq:us_conv_sigma} into the definition of $\sigma$ yields the Lippmann--Schwinger equation
\begin{equation}\label{eq:ls_sigma}
\sigma(\boldsymbol x) - k^2 V(\boldsymbol x)\int_{\Omega} G(\boldsymbol x- \boldsymbol y)\,\sigma(\boldsymbol y)\,d\boldsymbol y
=
k^2 V(\boldsymbol x)\,u_i(\boldsymbol x).
\end{equation}
Once $\sigma$ is obtained, the scattered field on $\mathbb{R}^d$ can be evaluated via \eqref{eq:us_conv_sigma}.

Since $V$ is compactly supported, we restrict computations to a box
$
\Omega = [-L,L]^d
$
containing $\mathrm{supp}(V)$ and discretize it by a uniform Cartesian grid.
Let $\{x_j\}$ be the grid points and denote $\sigma_j=\sigma(\boldsymbol x_j)$, $V_j=V(\boldsymbol x_j)$.
Discretization of \eqref{eq:ls_sigma} leads to the linear system
\begin{equation}\label{eq:disc_ls_sigma}
\sigma_j - k^2 V_j \sum_{k} G(\boldsymbol x_j-\boldsymbol x_k)\,\sigma_k
=
k^2 V_j\,u_i(\boldsymbol x_j).
\end{equation}
The discrete integral operator in \eqref{eq:disc_ls_sigma} has a convolution structure, and the associated matrix is Toeplitz. 
Therefore, matrix--vector products can be evaluated efficiently using FFT-based convolution.
The linear system \eqref{eq:disc_ls_sigma} is then solved using GMRES.

\subsection{Inverse Solver}

\subsubsection{Efficient computation of the operators $K_n$}

Recall that the operator $K_n$ is defined by
\[
K_n(V_1,\dots,V_n)
=
\sum_{j=0}^{n}
u_j(V_1,\dots,V_j)\,
u^*_{n-j}(V_{j+1},\dots,V_n),
\]
where
\[
u_j(V_1,\dots,V_j)(\boldsymbol x)
=
\int_{\Omega} G(\boldsymbol x,\boldsymbol y_1)V_1(\boldsymbol y_1)
\int_{\Omega} G(\boldsymbol y_1,\boldsymbol y_2)V_2(\boldsymbol y_2)\cdots
\int_{\Omega} G(\boldsymbol y_{j-1},\boldsymbol y_j)V_j(\boldsymbol y_j)u_0(\boldsymbol y_j)\,d\boldsymbol y_1\cdots d\boldsymbol y_j .
\]

\medskip

A direct evaluation of these nested integrals is computationally expensive.
In particular, computing $u_j(V_1,\dots,V_j)$ independently for each $j$ requires on the order of
\[
1+2+\cdots+n = \mathcal{O}(n^2)
\]
convolution operations.
To reduce this cost, we exploit the recursive structure of the Born terms and reuse intermediate results.
Although $u_j(V_1,\dots,V_j)$ is written with potentials ordered from $V_1$ to $V_j$, the numerical evaluation proceeds naturally from the innermost layer outward. 
We therefore introduce a prefix recursion:
\[
P_0=u_0,\quad
P_k(\boldsymbol x)=\int_\Omega G(\boldsymbol x, \boldsymbol y)\,V_k(\boldsymbol y)\,P_{k-1}(\boldsymbol y)\,d\boldsymbol y,\quad k=1,\dots,n.
\]
It follows that
\[
P_k = u_k(V_k,V_{k-1},\dots,V_1).
\]
That is, $P_k$ corresponds to the $k$th Born term with the reverse ordering of the potentials.
Similarly, we define a suffix recursion:
\[
T_n=u_0,\quad
T_k(\boldsymbol x)=\int_{\Omega} G(\boldsymbol x, \boldsymbol y)\,V_{k+1}(\boldsymbol y)\,T_{k+1}(\boldsymbol y)\,d\boldsymbol y,\quad k=n-1,\dots,0,
\]
which satisfies
\[
T_k = u_{n-k}(V_{k+1},\dots,V_n).
\]
With these intermediate fields, the operator $K_n$ can be evaluated as
\[
K_n(V_1,\dots,V_n)=\sum_{j=0}^{n} P_j\,T_j^*.
\]
We note that the use of reversed prefix recursion does not affect the inverse Born series. 
Indeed, the IBS involves a summation over all compositions $(i_1,\dots,i_\ell)$ of $m$,
\[
\mathcal{K}_m(\varphi)
=
-
\sum_{\ell=2}^{m}
\sum_{i_1+\cdots+i_\ell=m}
\mathcal{K}_1
K_\ell\big(
\mathcal{K}_{i_1}(\varphi),\dots,\mathcal{K}_{i_\ell}(\varphi)
\big).
\]
Whenever a composition $(i_1,\dots,i_j,i_{j+1},\dots,i_\ell)$ appears in this summation, the reversed ordering $(i_j,\dots,i_1,i_{j+1},\dots,i_\ell)$ also appears.
Thus, evaluating the Born terms in reversed order amounts to a reorganization of the same set of contributions, while enabling the reuse of intermediate fields.
Consequently, after computing $\{P_k\}$ and $\{T_k\}$ once, evaluating $K_n$ requires only $2n$ convolutions, rather than $\mathcal{O}(n^2)$.
Therefore, the complexity of evaluating $K_n$ is reduced from $\mathcal{O}(n^2)$ to $\mathcal{O}(n)$ convolution operations.

\subsubsection{Recovering $V$ from nonuniform Fourier samples}\label{sub:Fourier}

In the weak scattering regime, the measured far-field data can be converted (via the corresponding linearized inverse map) into a collection of Fourier samples of the potential,
\[
\widehat V(\boldsymbol p_j)\approx \phi_j,\quad \boldsymbol p_j\in P\subset\mathbb{R}^d,\ \ j=1,\dots,|P|,
\]
where $P=\{\boldsymbol p_j\}$ is a nonuniform set of frequencies contained in a ball such as $|p|\le 2k$. 
To reconstruct $V$ on a spatial grid, let
\[
X=\{x_\ell\}_{\ell=1}^{|X|}\subset\mathbb{R}^d
\]
be the set of recovery locations, the uniform grid points in the computational domain. 
We discretize $V$ as
\[
v=\bigl(V(\boldsymbol x_\ell)\bigr)_{\ell=1}^{|X|
}\in\mathbb{C}^{|X|},
\quad 
\phi=\bigl(\phi_j\bigr)_{j=1}^{|P|}\in\mathbb{C}^{|P|}.
\]
Define the nonuniform Fourier transform (NUFFT) operator $\mathcal{F}_{X,P}:\mathbb{C}^{|X|}\to\mathbb{C}^{|P|}$ by
\begin{equation}\label{eq:nufft_forward}
\bigl(\mathcal{F}_{X,P}(v)\bigr)_j
=\sum_{\ell=1}^{|X|} v_\ell\,e^{-\,\mathrm{i}\,\boldsymbol x_\ell\cdot \boldsymbol p_j}, 
\quad j=1,\dots,|P|.
\end{equation}
Its adjoint  $\mathcal{F}_{P,X}^{*}:\mathbb{C}^{|P|}\to\mathbb{C}^{|X|}$ is defined by
\begin{equation}\label{eq:nufft_adjoint}
\bigl(\mathcal{F}_{P,X}^{*}(\phi)\bigr)_\ell
=\sum_{j=1}^{|P|} \phi_j\,e^{\,\mathrm{i}\,\boldsymbol x_\ell\cdot \boldsymbol p_j},
\quad \ell=1,\dots,|X|.
\end{equation}
Note that $\mathcal{F}_{P,X}^{*}$ is the adjoint of $\mathcal{F}_{X,P}$ and is not, in general, its inverse, since the sampling set $P$ is nonuniform and the system is typically overdetermined.

We recover $v$ by solving the least-squares problem
\begin{equation}\label{eq:lsq}
\min_{v\in\mathbb{C}^{|X|}} \ \|\mathcal{F}_{X,P}(v)-\phi\|_2^2,
\end{equation}
optionally with Tikhonov regularization,
\begin{equation}\label{eq:tikhonov}
\min_{v\in\mathbb{C}^{|X|}} \ \|\mathcal{F}_{X,P}(v)-\phi\|_2^2+\lambda^2\|v\|_2^2.
\end{equation}
The normal equations associated with \eqref{eq:tikhonov} are
\begin{equation}\label{eq:normal}
\bigl(\mathcal{F}_{P,X}^{*}\mathcal{F}_{X,P}+\lambda^2 I\bigr)v
=\mathcal{F}_{P,X}^{*}\phi.
\end{equation}
We solve \eqref{eq:normal} iteratively using GMRES. The iteration is terminated upon
convergence of the residual or when the prescribed maximum number of
iterations is reached. In our implementation, the restart parameter is set
to $200$ and the maximum number of iterations is set to $100$. 
Each application of the normal operator
\[
v\ \mapsto\ \mathcal{F}_{P,X}^{*}\bigl(\mathcal{F}_{X,P}(v)\bigr)
\]
is evaluated efficiently by two NUFFT calls: one forward NUFFT for $\mathcal{F}_{X,P}$ and one adjoint NUFFT for $\mathcal{F}_{P,X}^{*}$. 
Thus, one GMRES iteration costs $\mathcal{O}((|X|+|P|)\log(|X|+|P|))$ , rather than $\mathcal{O}(|X||P|)$ for direct evaluation.
\subsubsection{Comparison of numerical algorithms}

We summarize the reconstruction procedures for the three types of measurement data considered in this work. In all cases, the reconstruction of phaseless data is performed using the IBS, while the main differences for the problems considered lie in how the linear operator $\mathcal{K}_1$ is constructed from the measurements.

When phase information is available, the standard complex-valued Born series defined in \eqref{def of original born} is used. 
In this situation, there is no essential difference between measuring the total field and the scattered field. 
For general measurements, $\mathcal{K}_1^p$ is implemented as a regularized pseudoinverse of the first-order Born operator $K_1^p$, defined in \eqref{def of phase born} and \eqref{Forward of phase}, similar to \cite{defilippisBornInverseBorn2023}. 
For far-field measurements, the data are first converted into Fourier samples of the potential using the far-field approximation (Section~\ref{sub:farfield}), and the potential is reconstructed through the NUFFT-based least-squares inversion described in Section~\ref{sub:Fourier}.

When only phaseless measurements of the total field are available, we employ the phaseless Born series \eqref{def of born}. In this case the data are first converted into the quantity $\psi = |u|^2 - |u_0|^2$, from which the linear operator $K_1$ is constructed. 
For general data, a numerical pseudoinverse of $K_1$ is used to obtain $\mathcal{K}_1$. 
For far-field data, we first discard Fourier samples
\(\widehat{V}(k(\hat{\boldsymbol x}-\hat{\boldsymbol y}))\), referred to as
``bad'' samples, whenever the linear systems in
\eqref{component 1} and \eqref{component 2} are ill-conditioned.
The filtering criterion is based on the condition number of the associated
\(2\times 2\) system: samples are discarded if this condition number exceeds
a prescribed threshold. Since the sum of the squared singular values of this
matrix is fixed, ill-conditioning occurs precisely when the smaller singular
value becomes close to zero. In practice, we therefore discard direction pairs
for which the smaller singular value is below \(10^{-2}\). 
Then, the Fourier method described in Section~\ref{sub:fourier method} is used to recover the Fourier coefficients of the potential, which are subsequently inverted via the NUFFT-based reconstruction in Section~\ref{sub:Fourier}.

Finally, when only phaseless measurements of the scattered field are available, we first apply the polarization method (Section~\ref{sub:polar}) to construct an approximate potential $V^a$. 
This approximation is then used to infer an approximate phase of the scattered data. 
The reconstructed complex-valued data are subsequently treated as phase data, and the standard IBS procedure described above is applied.
\begin{remark}
    Since the polarization method relies on the Born approximation, it is expected to be effective for low-contrast scattering potentials, while its accuracy deteriorates for high-contrast potentials. This behavior is consistent with the numerical results presented below.
\end{remark}
The complete reconstruction procedures are summarized below.

\begin{itemize}

\item \textbf{Algorithm 1: Data with phase}

\begin{enumerate}
\item Collect the complex-valued measurements.
\item Construct the first-order Born operator $K_1^p$.
\item 
\begin{itemize}
\item[{}] general data: compute a regularized pseudoinverse of $K_1^p$ to obtain $\mathcal{K}_1^p$.
\item[{}] Far-field data: convert the measurements to Fourier samples of $V$ using the far-field approximation and construct $\mathcal{K}_1$ via the NUFFT-based least-squares inversion.
\end{itemize}
\item Apply the inverse Born series
\[
V \approx \sum_{m=1}^{M}\mathcal{K}_m(u_s).
\]
\end{enumerate}

\item \textbf{Algorithm 2: Phaseless data of the total field}

\begin{enumerate}
\item Measure the phaseless total-field data $|u|$.
\item Compute the data difference
\[
\psi = |u|^2 - |u_0|^2.
\]
\item Construct the linear operator $K_1$ associated with the Born expansion.
\item 
\begin{itemize}
\item[{}] General data: compute a numerical pseudoinverse of $K_1$ to obtain $\mathcal{K}_1$.
\item[{}] Far-field data: Discard ``bad'' Fourier samples; recover the Fourier coefficients of $V$ via the Fourier method; and construct $\mathcal{K}_1$ via the NUFFT-based least-squares inversion.
\end{itemize}
\item Apply the inverse Born series using $\{\mathcal{K}_m\}$ to reconstruct $V$.
\end{enumerate}

\item \textbf{Algorithm 3: Phaseless data of the scattered field}

\begin{enumerate}
\item Measure the phaseless scattered-field data $|u_s|$.
\item Apply the polarization identity to obtain an approximate potential $V^a$.
\item Use $V^a$ to compute the scattered field $u_s^a$. 
Construct an approximate complex-valued dataset by combining the phase of $u_s^a$ with the measured magnitude $|u_s|$:
\[
u_s^{\mathrm{rec}}
=
\frac{u_s^a}{|u_s^a|}\,|u_s|.
\]
\item Treat $u_s^{\mathrm{rec}}$ as complex-valued measurements and apply Algorithm~1 to perform the IBS reconstruction.
\end{enumerate}

\end{itemize}


\section{Numerical simulations}\label{sec:result}

In this section, we present numerical results to verify our numerical algorithms. 
The computational domain is $\Omega = [-6.4, 6.4] \times [-6.4, 6.4]\subseteq \mathbb{R}^2$. 
The grid resolution is $128 \times 128$, with mesh sizes $d_x = 0.1$ and $d_y = 0.1$. In order to avoid ``inverse crime", we double the grid size when generating the forward data. 
The wavenumber is set to $k = 5.0$. 
We use 400 incident waves of the form $u_0(\boldsymbol x) = e^{\mathrm{i} \boldsymbol{k} \cdot \boldsymbol x}$, where $|\boldsymbol{k}| = k$ and the directions are uniformly distributed on the unit circle.

We consider two types of potentials in our numerical experiments.
The smoothed circular potential is defined as
\begin{equation*}
V( x,  y)
=
\frac{A}{2}
\left[
1 - \tanh\left(
\frac{\sqrt{( x-x_c)^2 + (y-y_c)^2} - R}{\delta}
\right)
\right],
\end{equation*}
where $( x_c,y_c)$ is the center of the circle, $R$ is the radius, $\delta$ controls the smoothness of the boundary, and $A$ is the amplitude. 
In our experiments, we take
\begin{equation*}
(x_c, y_c) = (0,0),  \quad
R = 2.55, \quad
\delta = 0.255.
\end{equation*}
The Gaussian mixture potential is given by
\begin{equation*}
V( x,  y)
=
A\sum_{i=1}^{2}\exp\!\left(
-\frac{(x-x_i)^2 + (y- y_i)^2}{2\sigma_i^2}
\right).
\end{equation*}
The parameters are
\begin{align*}
( x_1,y_1) &= (-1.905,\; 1.905), & \sigma_1 &= 1.524, \\
(x_2,y_2) &= ( 2.54,\; -2.54), & \sigma_2 &= 1.143.
\end{align*}

We consider different amplitudes $A$ for these two types of potentials to test the performance of our algorithms. 
In the following subsections, we present the numerical results. 
For each figure, we compare reconstructions obtained from phase and phaseless data. 
The relative errors are reported in the tables below each figure, computed as
\begin{equation*}
\frac{\|V - V_{\mathrm{true}}\|_2}{\|V_{\mathrm{true}}\|_2}.
\end{equation*}
We also compute the projection result and its error as a baseline, where 
$V_{\mathrm{proj}} = \mathcal{K}_1 K_1(d)$ and $d$ denotes the available data.
\subsection{Direct method}

For the experiments presented in this subsection, the detectors are equally spaced on the boundary $\partial \Omega$; their total number is 512, which matches the number of grid points on the boundary. 
The regularization parameters are chosen empirically for each data type.
Several candidate values were tested for each type of data, and the value
yielding stable reconstructions with relatively small reconstruction errors was
selected. The selected value was then fixed for all experiments with the same
data type. For the direct method, we use
\[
\lambda_{\mathrm{phase}}=2\times10^{-1},
\qquad
\lambda_{\mathrm{phaseless}}=4\times10^{-2}.
\]

\begin{figure}[htbp]
\centering

\begin{subfigure}[b]{0.95\textwidth}
\includegraphics[width=\textwidth]{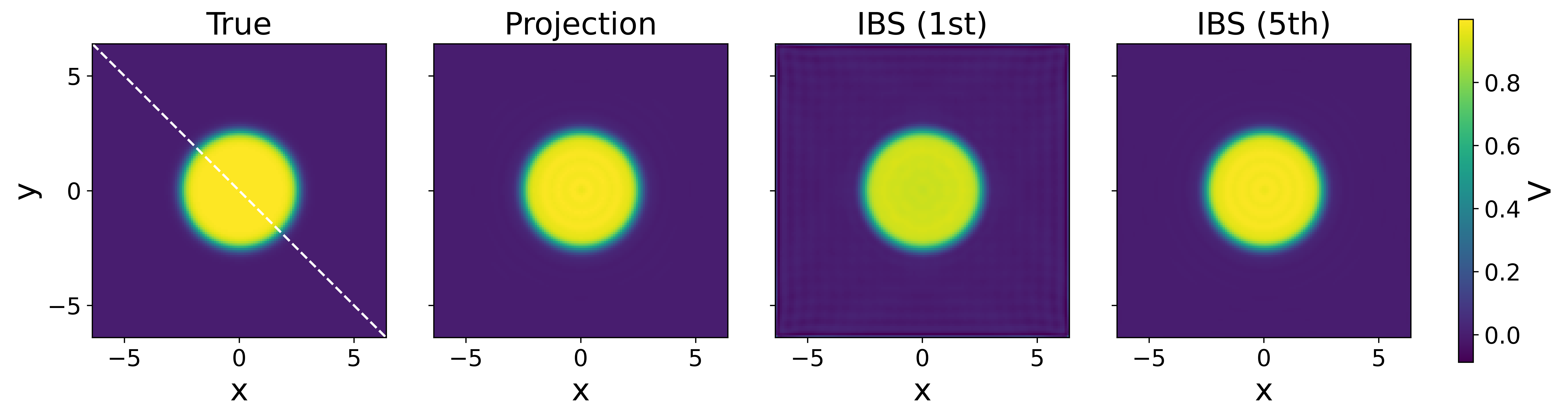}
\caption{Reconstruction of $V$ using phase data}
\end{subfigure}

\begin{subfigure}[b]{0.47\textwidth}
\includegraphics[width=\textwidth]{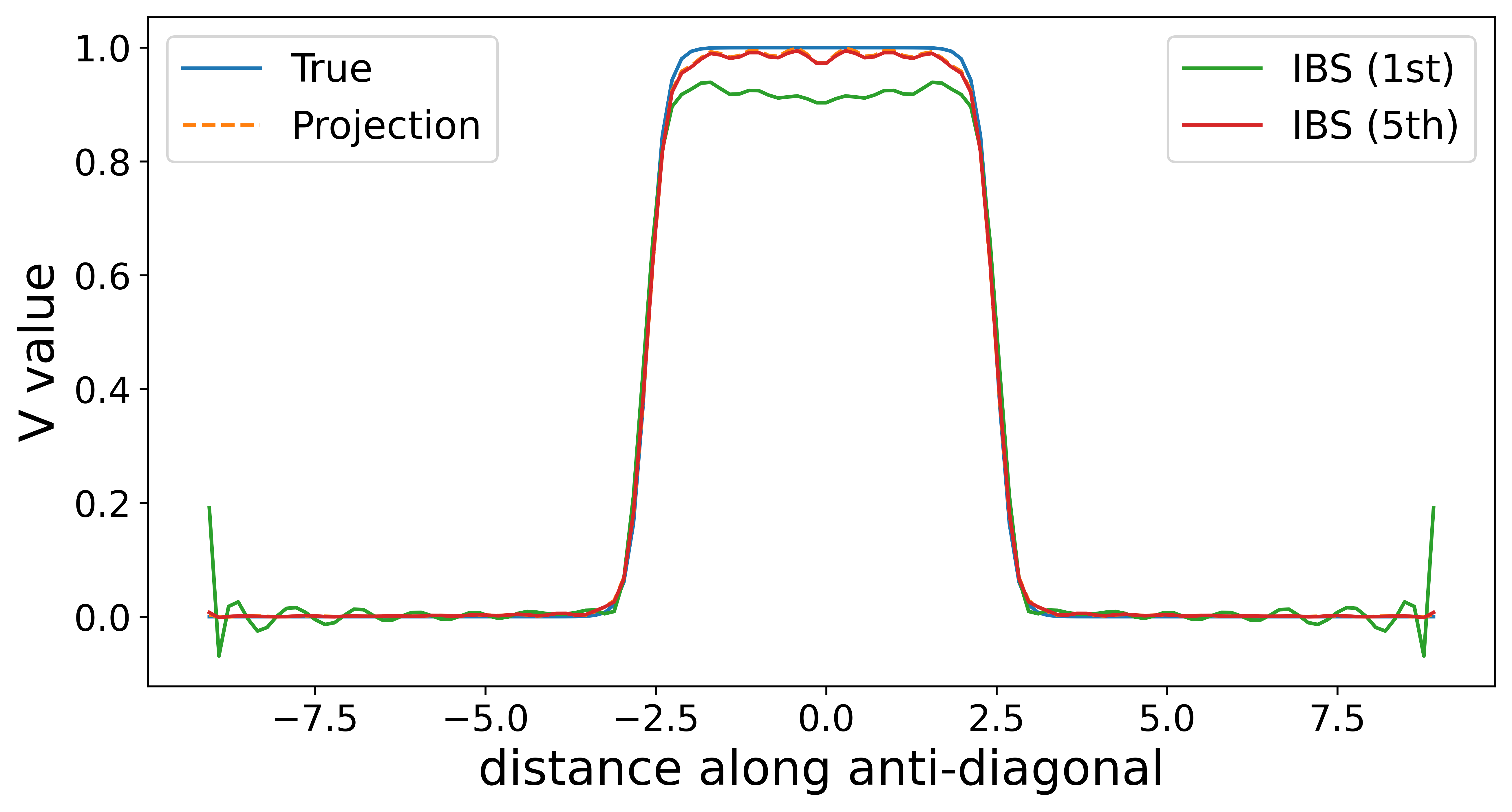}
\caption{Anti-diagonal cross section}
\end{subfigure}

\begin{subfigure}[b]{0.95\textwidth}
\includegraphics[width=\textwidth]{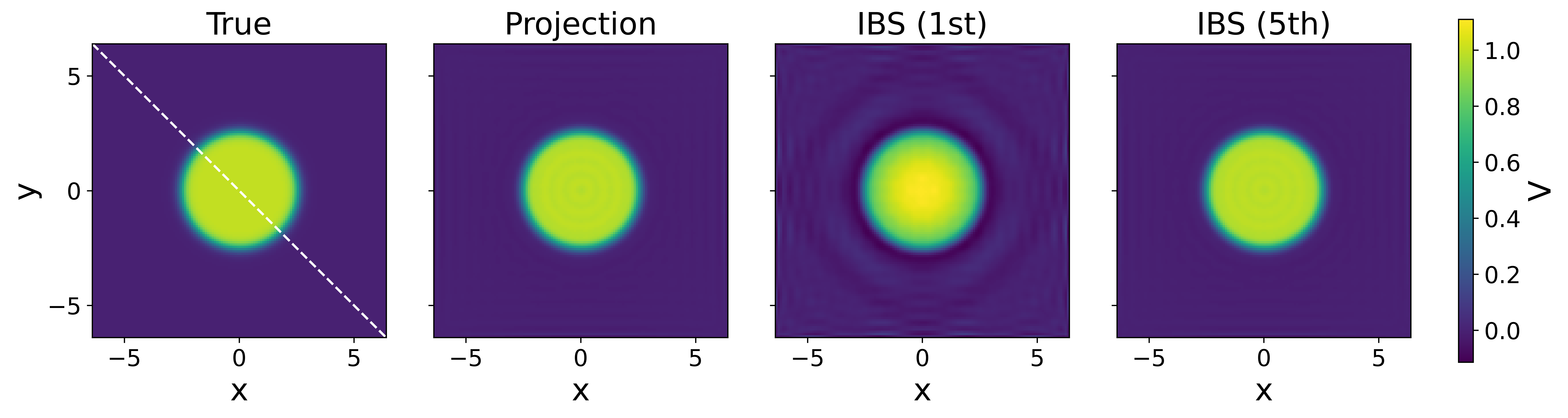}
\caption{Reconstruction of $V$ using phaseless data}
\end{subfigure}

\begin{subfigure}[b]{0.47\textwidth}
\includegraphics[width=\textwidth]{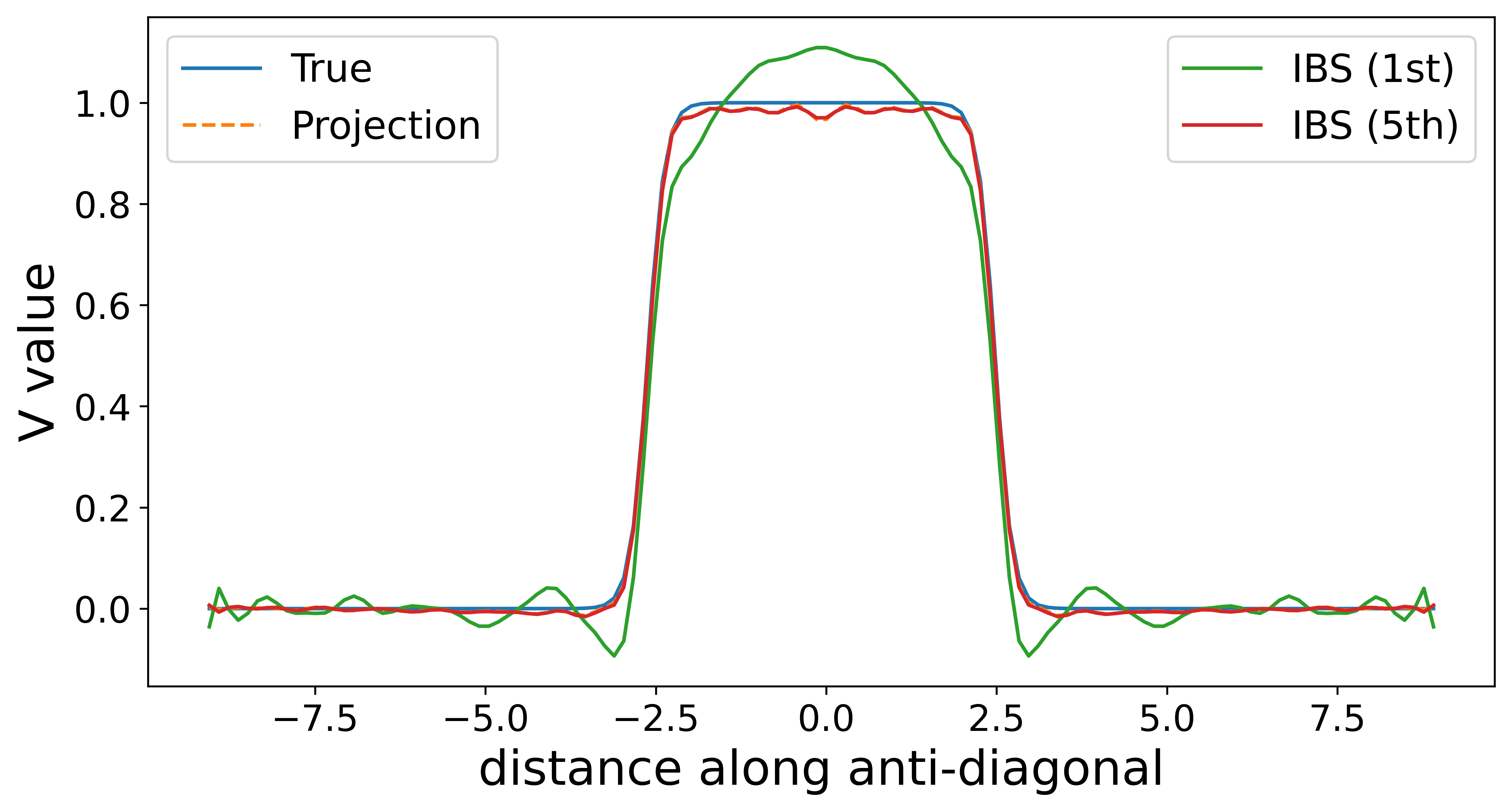}
\caption{Anti-diagonal cross section}
\end{subfigure}

\caption{Reconstructions of the low-contrast disk using the direct method.}
\label{fig:direct-good-circle}
\end{figure}

\begin{figure}[htbp]

\centering
\begin{subfigure}[b]{0.95\textwidth}
\includegraphics[width=\textwidth]{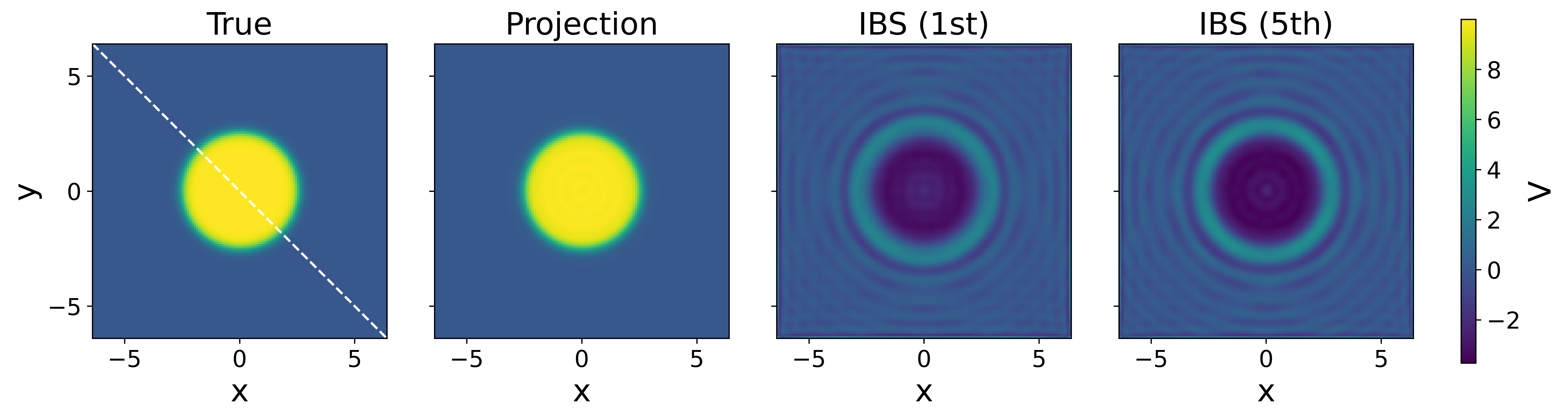}
\caption{Reconstruction of $V$ using phase data }
\end{subfigure}

\begin{subfigure}[b]{0.50\textwidth}
\includegraphics[width=\textwidth]{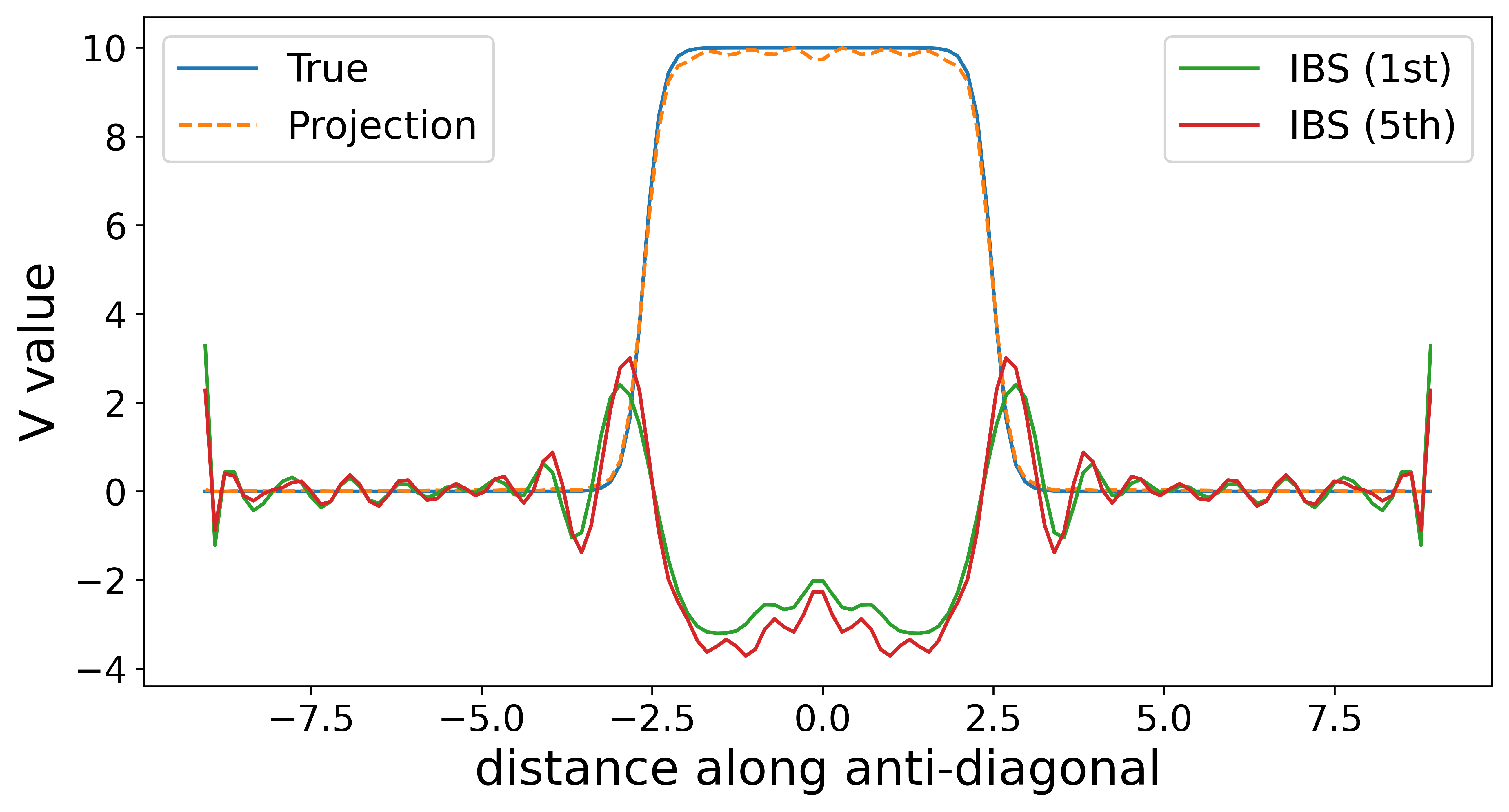}
\caption{Anti-diagonal cross section }
\end{subfigure}\hfill
\begin{subfigure}[b]{0.95\textwidth}
\includegraphics[width=\textwidth]{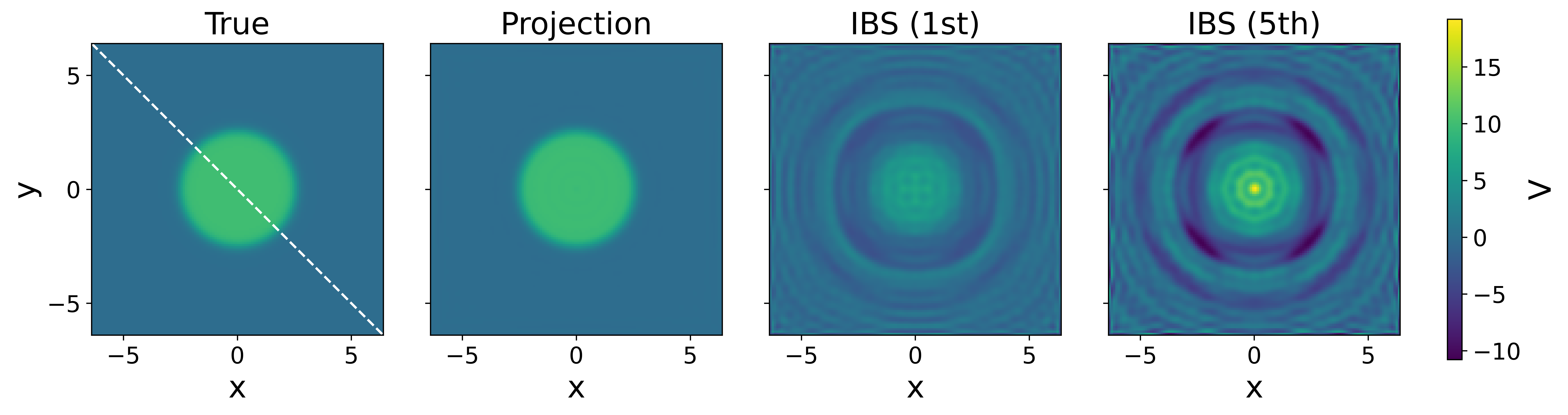}
\caption{Reconstruction of $V$ using phaseless data }
\end{subfigure}

\begin{subfigure}[b]{0.50\textwidth}
\includegraphics[width=\textwidth]{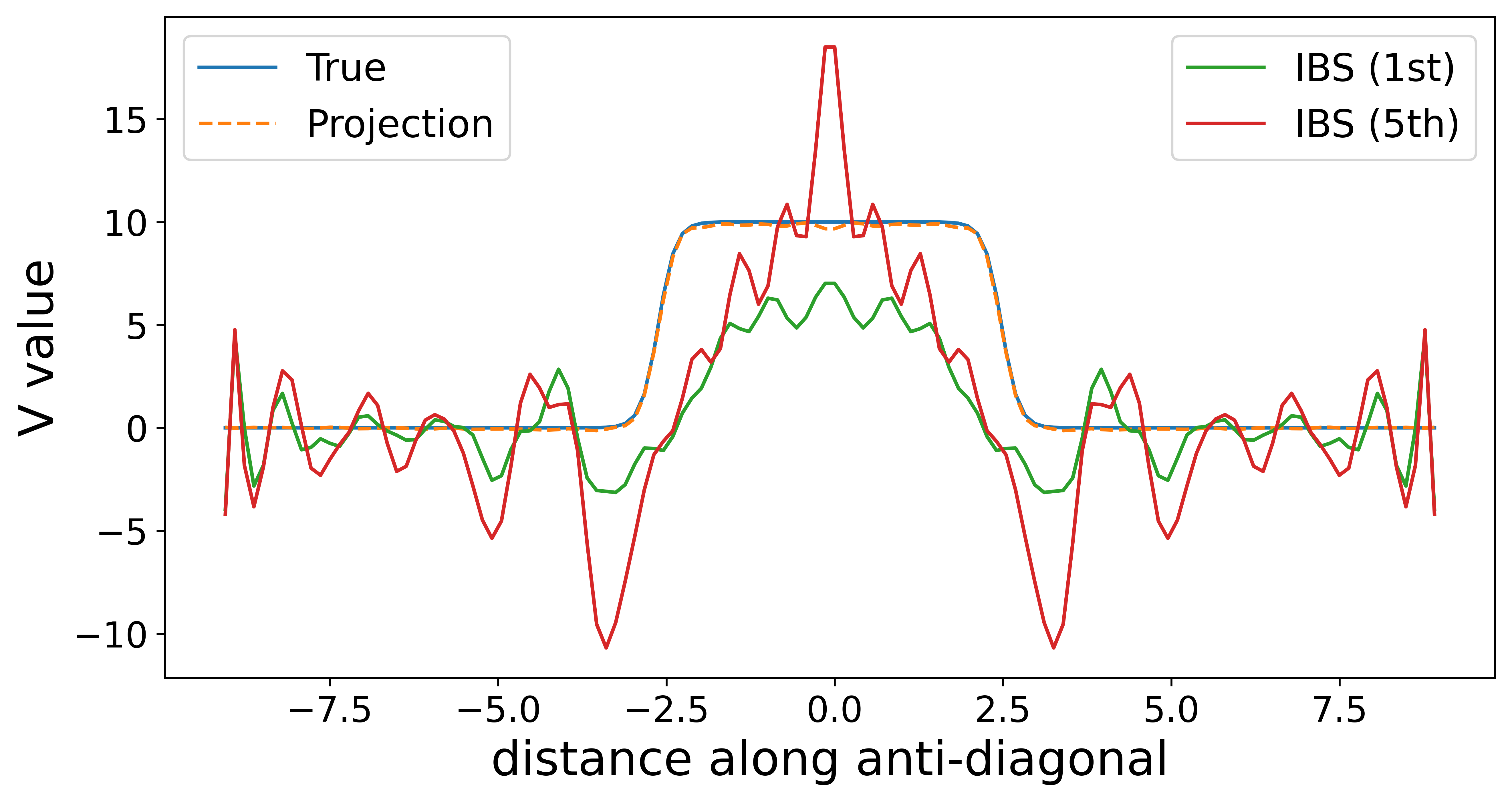}
\caption{Anti-diagonal cross section}
\end{subfigure}

\caption{Reconstructions of high contrast disk using direct method}
\label{fig:direct-bad-circle}
\end{figure}

\begin{figure}[htbp]
\centering

\begin{subfigure}[b]{0.95\textwidth}
\includegraphics[width=\textwidth]{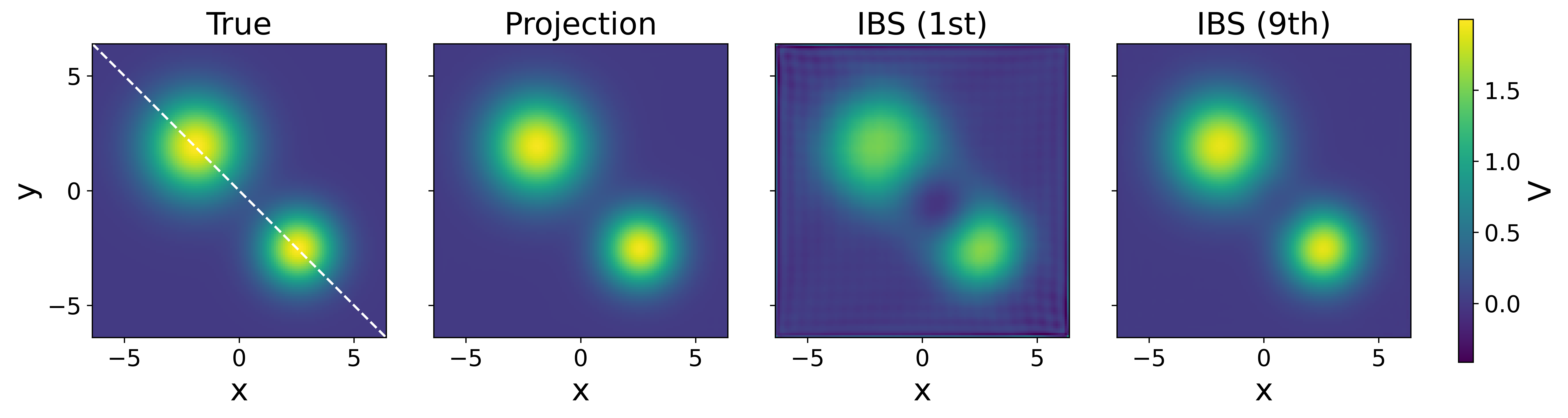}
\caption{Reconstruction of $V$ using phase data}
\end{subfigure}

\begin{subfigure}[b]{0.45\textwidth}
\includegraphics[width=\textwidth]{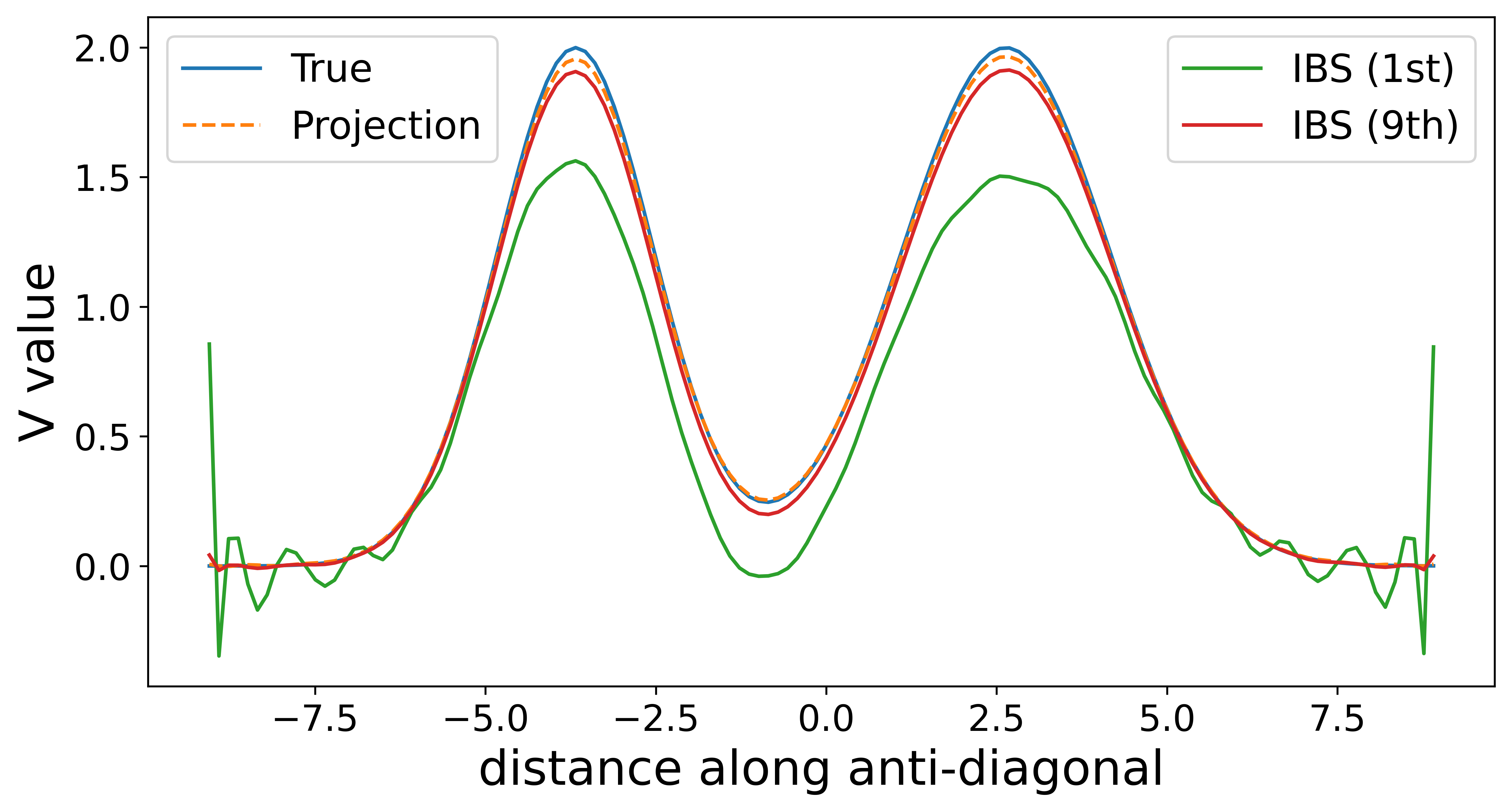}
\caption{Anti-diagonal cross section}
\end{subfigure}

\begin{subfigure}[b]{0.95\textwidth}
\includegraphics[width=\textwidth]{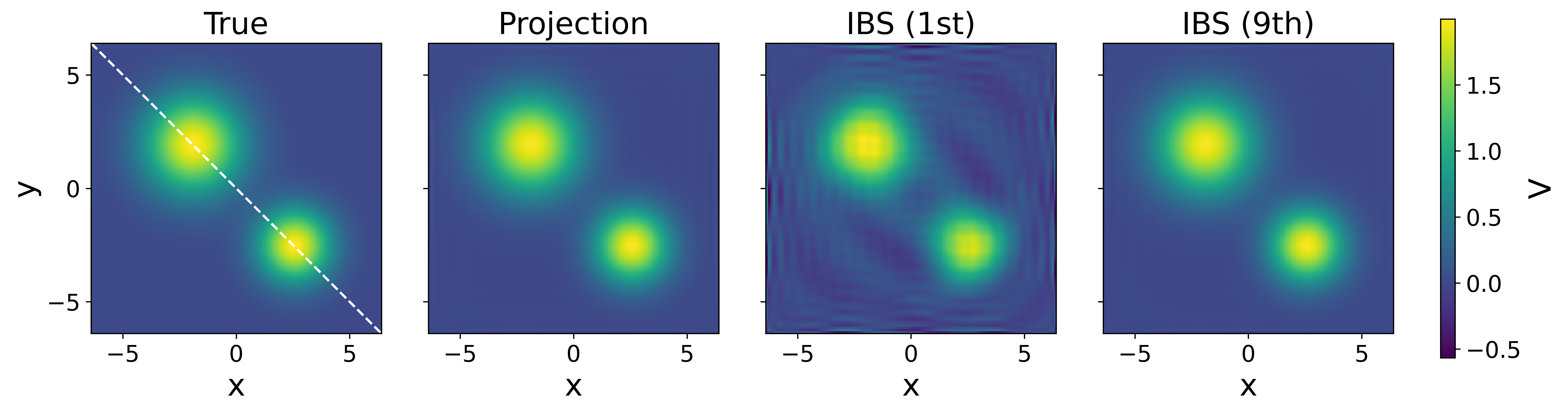}
\caption{Reconstruction of $V$ using phaseless data}
\end{subfigure}

\begin{subfigure}[b]{0.45\textwidth}
\includegraphics[width=\textwidth]{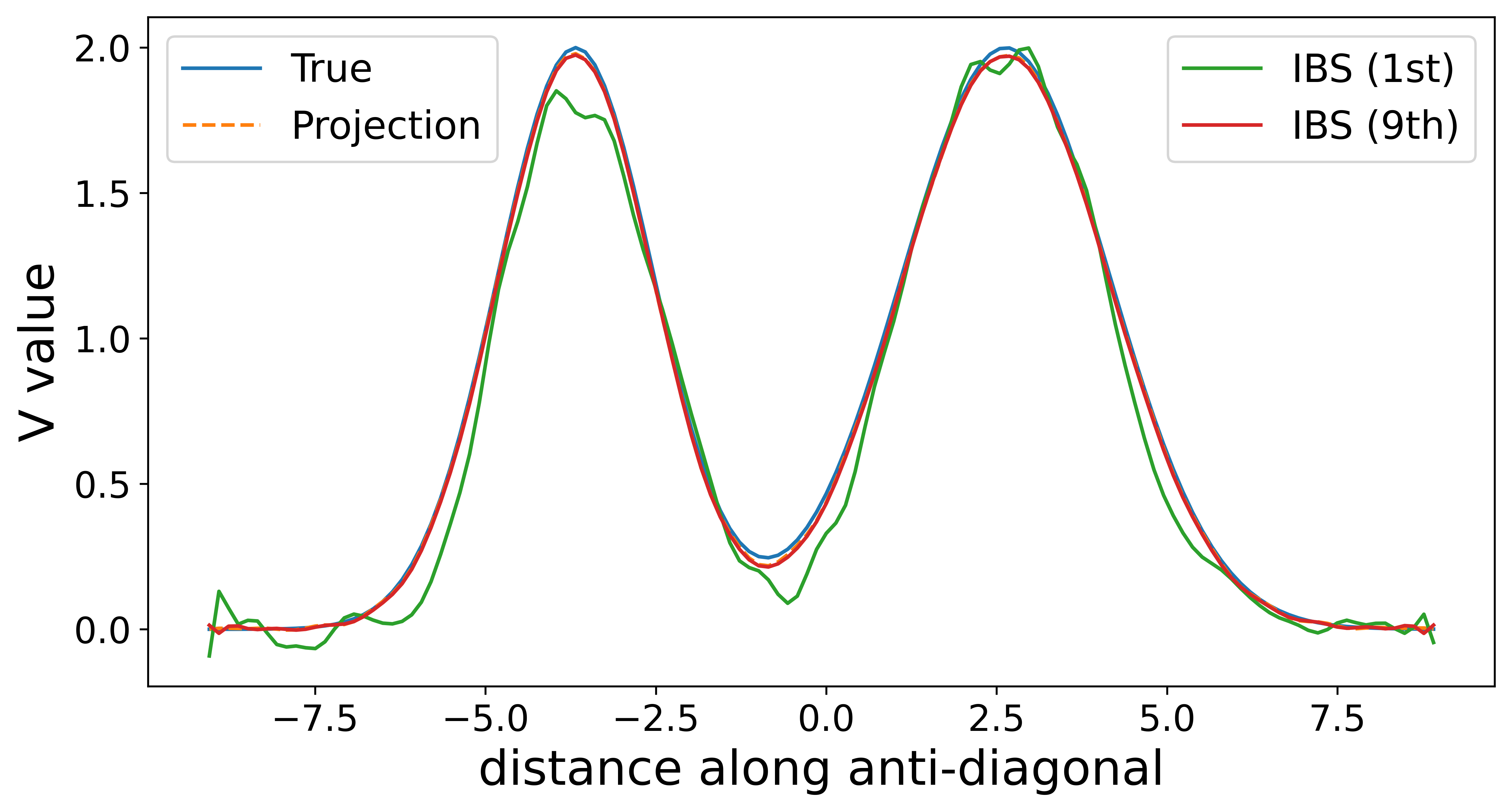}
\caption{Anti-diagonal cross section}
\end{subfigure}

\caption{Reconstructions of the low-contrast Gaussian mixture using the direct method.}
\label{fig:direct-good-gaussian}
\end{figure}

\begin{figure}[htbp]

\centering
\begin{subfigure}[b]{0.95\textwidth}
\includegraphics[width=\textwidth]{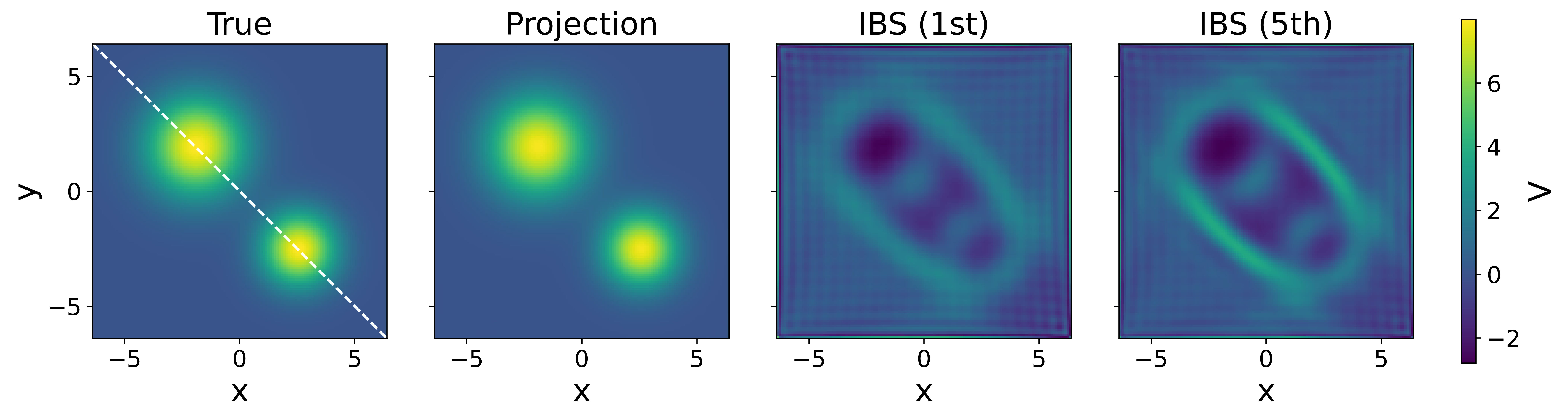}
\caption{Reconstruction of $V$ using phase data }
\end{subfigure}

\begin{subfigure}[b]{0.50\textwidth}
\includegraphics[width=\textwidth]{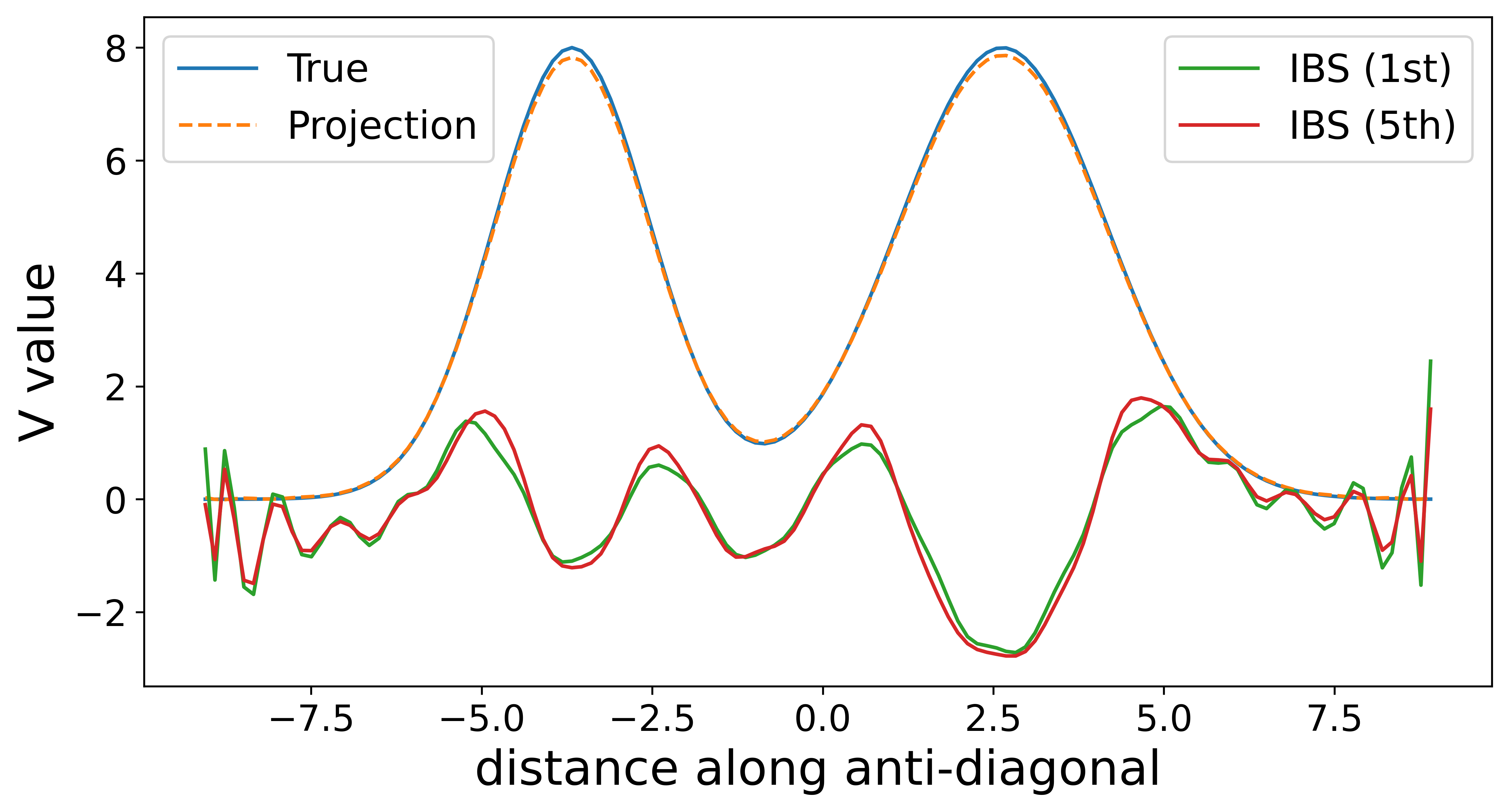}
\caption{Anti-diagonal cross section }
\end{subfigure}\hfill
\begin{subfigure}[b]{0.95\textwidth}
\includegraphics[width=\textwidth]{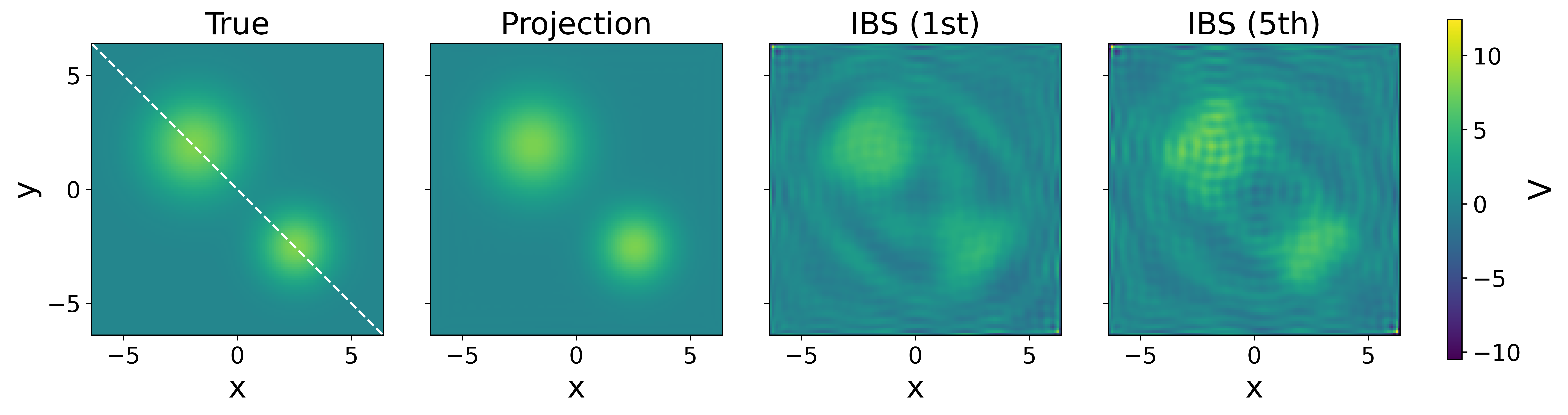}
\caption{Reconstruction of $V$ using phaseless data }
\end{subfigure}

\begin{subfigure}[b]{0.50\textwidth}
\includegraphics[width=\textwidth]{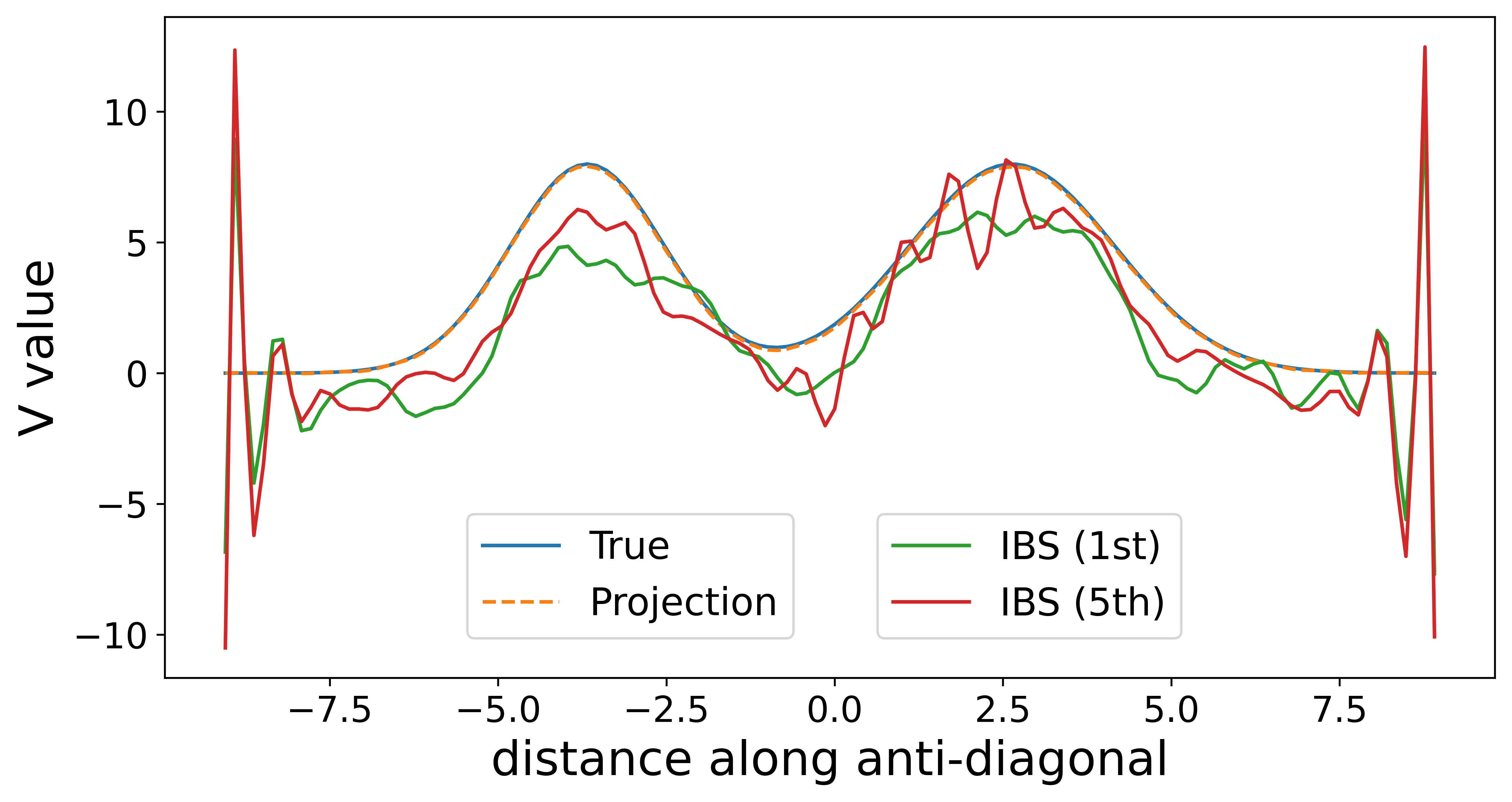}
\caption{Anti-diagonal cross section}
\end{subfigure}

\caption{Reconstructions of high contrast Gaussian mixture using direct method}
\label{fig:direct-bad-gaussian}
\end{figure}

In Figures \ref{fig:direct-good-circle} and \ref{fig:direct-bad-circle}, we present the numerical results for low- and high-contrast disk potentials, with amplitudes $A = 1.0$ and $A = 10.0$, respectively. 
In Figures \ref{fig:direct-good-gaussian} and \ref{fig:direct-bad-gaussian}, we present the numerical results for low- and high-contrast Gaussian mixture potentials, with amplitudes $A = 2.0$ and $A = 8.0$, respectively. 

As expected from the theoretical analysis, when the contrast is low, the method works well, and the reconstructions from phase data are slightly better than those from phaseless data. 
When the contrast is high, the IBS iteration fails to converge in both cases. 
The largest admissible IBS order is indicated in the corresponding figures.

\begin{table}[htbp]
\centering
\caption{Relative reconstruction errors corresponding to
Fig.~\ref{fig:direct-good-circle}.}
\label{tab:direct-good-circle}
\small
\renewcommand{\arraystretch}{1.15}
\setlength{\tabcolsep}{5pt}
\begin{tabular}{lcccccc}
\toprule
 & Projection & IBS1 & IBS2 & IBS3 & IBS4 & IBS5 \\
\midrule
Phase data
& 0.0220 & 0.0950 & 0.0539 & 0.0256 & 0.0251 & 0.0233 \\
Phaseless data
& 0.0304 & 0.1394 & 0.0383 & 0.0344 & 0.0329 & 0.0330 \\
\bottomrule
\end{tabular}
\end{table}

\begin{table}[htbp]
\centering
\caption{Relative reconstruction errors corresponding to
Fig.~\ref{fig:direct-good-gaussian}.}
\label{tab:direct-good-gaussian}
\small
\renewcommand{\arraystretch}{1.15}
\setlength{\tabcolsep}{5pt}
\begin{tabular}{lcccccc}
\toprule
 & Projection & IBS1 & IBS3 & IBS5 & IBS7 & IBS9 \\
\midrule
Phase data
& 0.0147 & 0.2289 & 0.0864 & 0.0524 & 0.0382 & 0.0314 \\
Phaseless data
& 0.0359 & 0.2020 & 0.0431 & 0.0392 & 0.0387 & 0.0386 \\
\bottomrule
\end{tabular}
\end{table}

\subsection{Fourier method}\label{sec:fourier}

In this subsection, we consider far-field data. 
The detectors are equally distributed on a circle centered at the origin with radius $R = 300$, in the same manner as the incident directions. 
The regularization parameter $\lambda$ is chosen as follows: $\lambda = 10$ for phase data, $\lambda = 20$ for phaseless data with disk-type potentials, and $\lambda = 30$ for phaseless data with Gaussian-type potentials. 
In all our tests, we discard $\approx 12\%$ of the Fourier samples in order to avoid solving ill-posed linear systems.

\begin{figure}[htbp]
\centering

\begin{subfigure}[b]{0.95\textwidth}
\includegraphics[width=\textwidth]{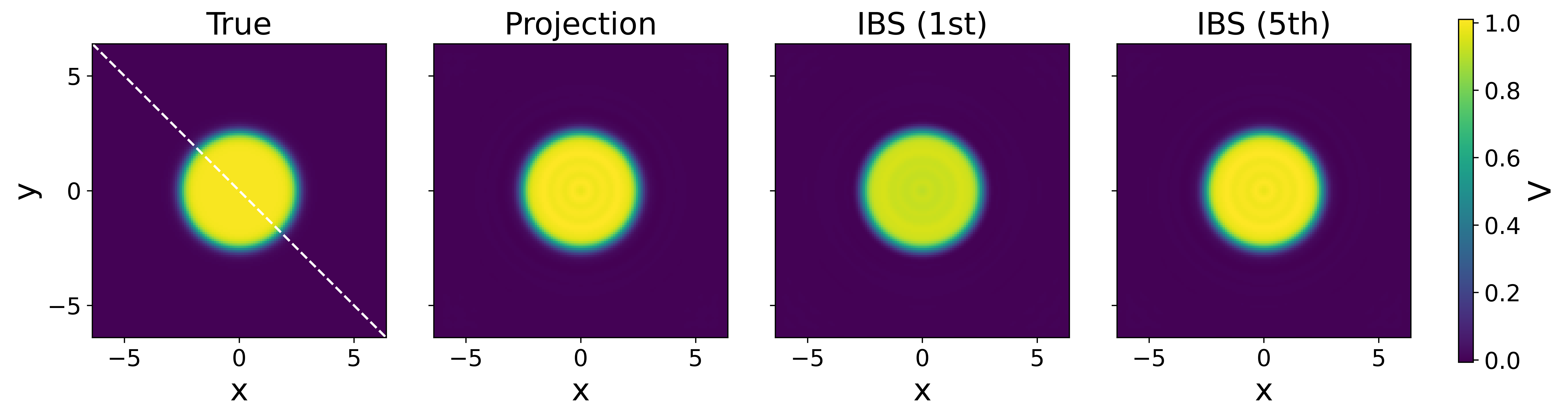}
\caption{Reconstruction of $V$ using phase data}
\end{subfigure}

\begin{subfigure}[b]{0.50\textwidth}
\includegraphics[width=\textwidth]{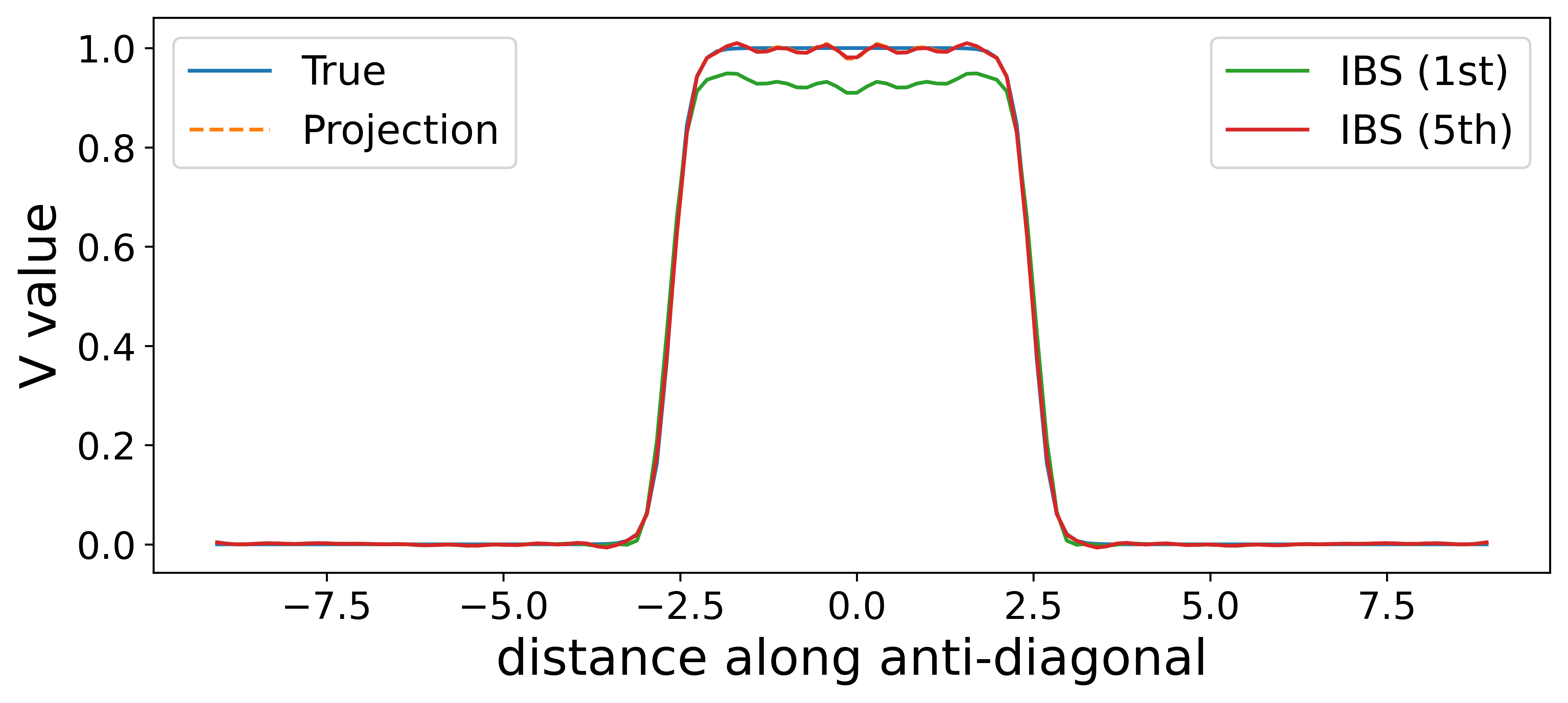}
\caption{Anti-diagonal cross section}
\end{subfigure}

\begin{subfigure}[b]{0.95\textwidth}
\includegraphics[width=\textwidth]{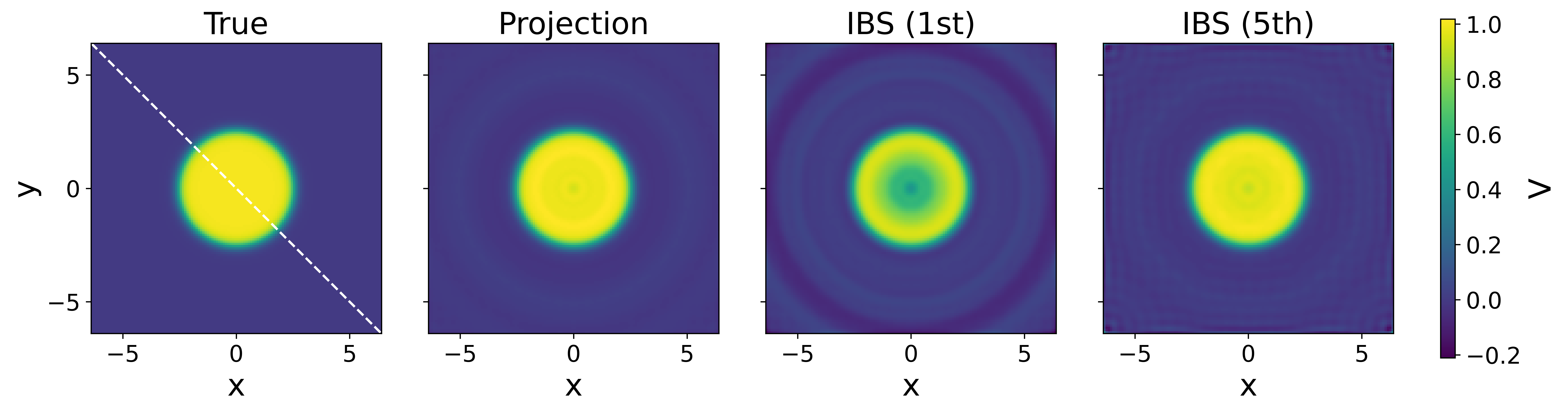}
\caption{Reconstruction of $V$ using phaseless data}
\end{subfigure}

\begin{subfigure}[b]{0.50\textwidth}
\includegraphics[width=\textwidth]{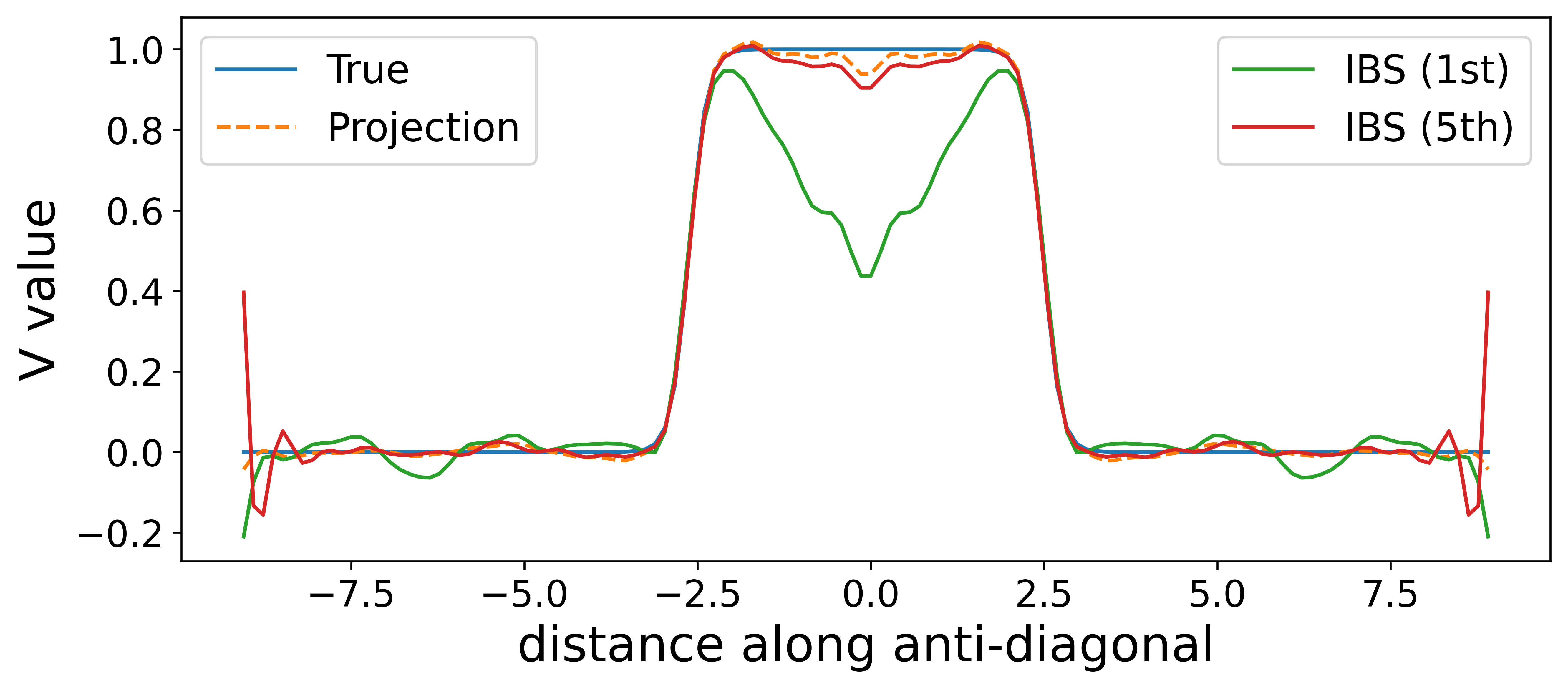}
\caption{Anti-diagonal cross section}
\end{subfigure}

\caption{Reconstructions of the low-contrast disk using the Fourier method.}
\label{fig:fourier-good-circle}
\end{figure}

\begin{figure}[htbp]
\centering

\begin{subfigure}[b]{0.95\textwidth}
\includegraphics[width=\textwidth]{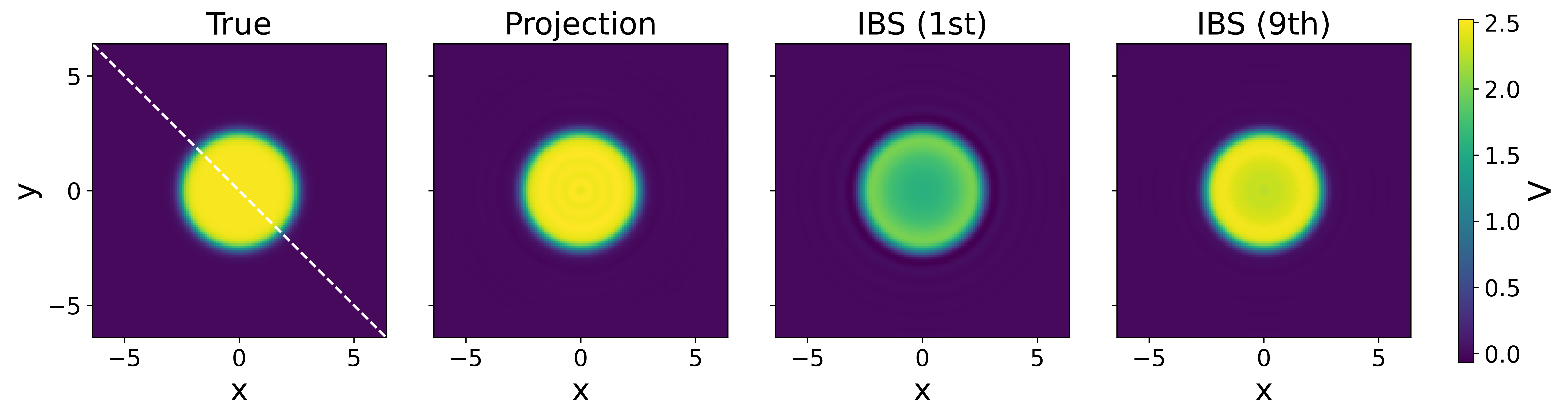}
\caption{Reconstruction of $V$ using phase data}
\end{subfigure}

\begin{subfigure}[b]{0.50\textwidth}
\includegraphics[width=\textwidth]{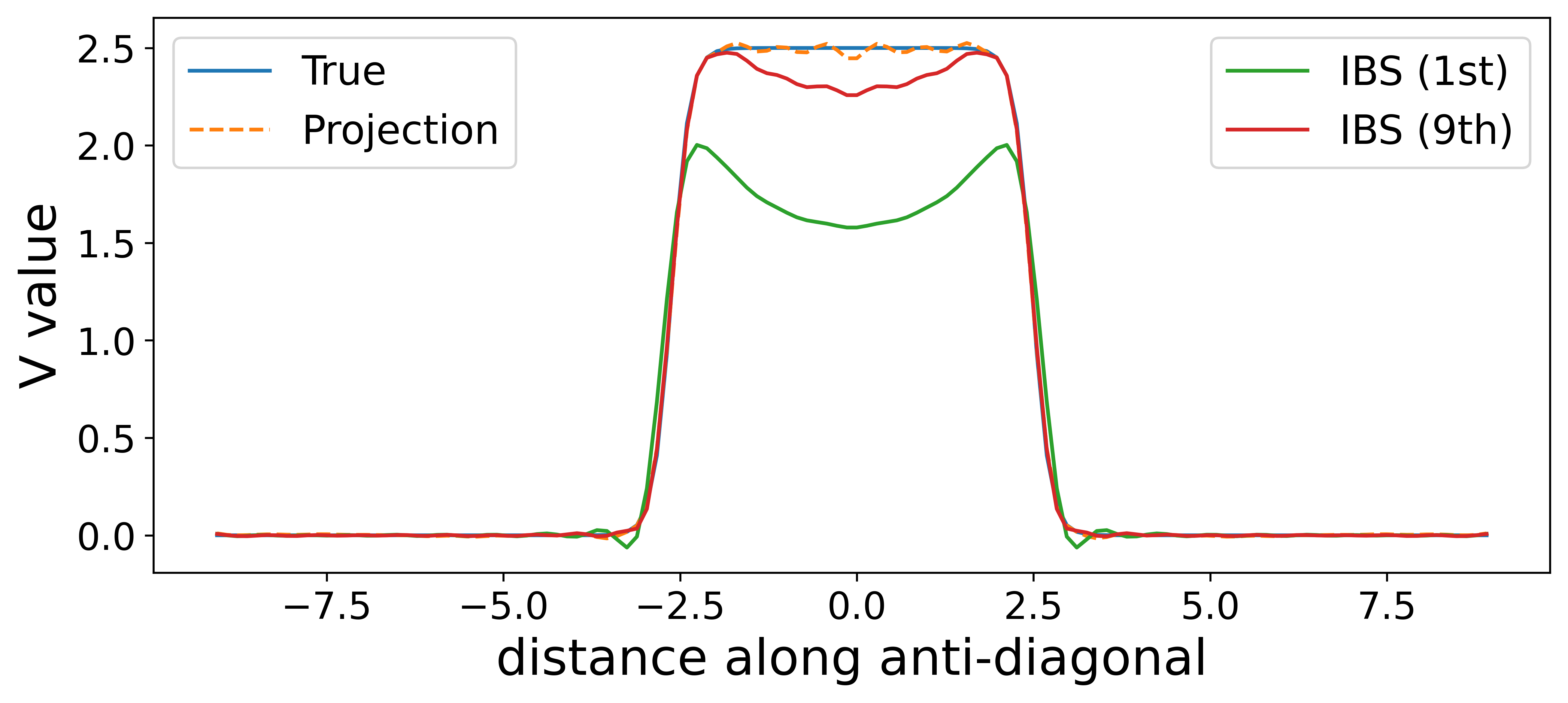}
\caption{Anti-diagonal cross section}
\end{subfigure}

\begin{subfigure}[b]{0.95\textwidth}
\includegraphics[width=\textwidth]{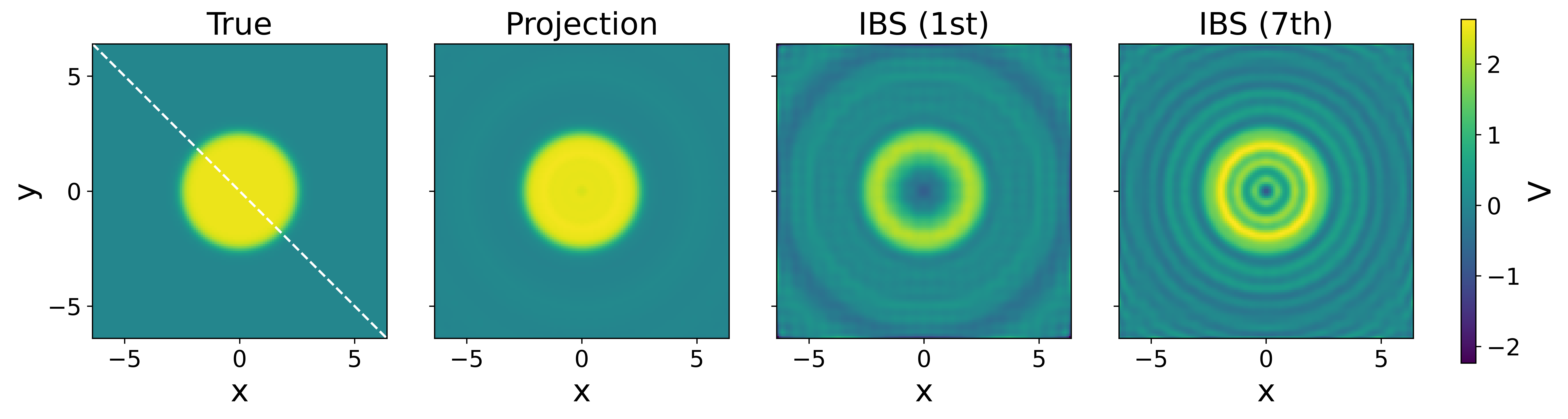}
\caption{Reconstruction of $V$ using phaseless data}
\end{subfigure}

\begin{subfigure}[b]{0.50\textwidth}
\includegraphics[width=\textwidth]{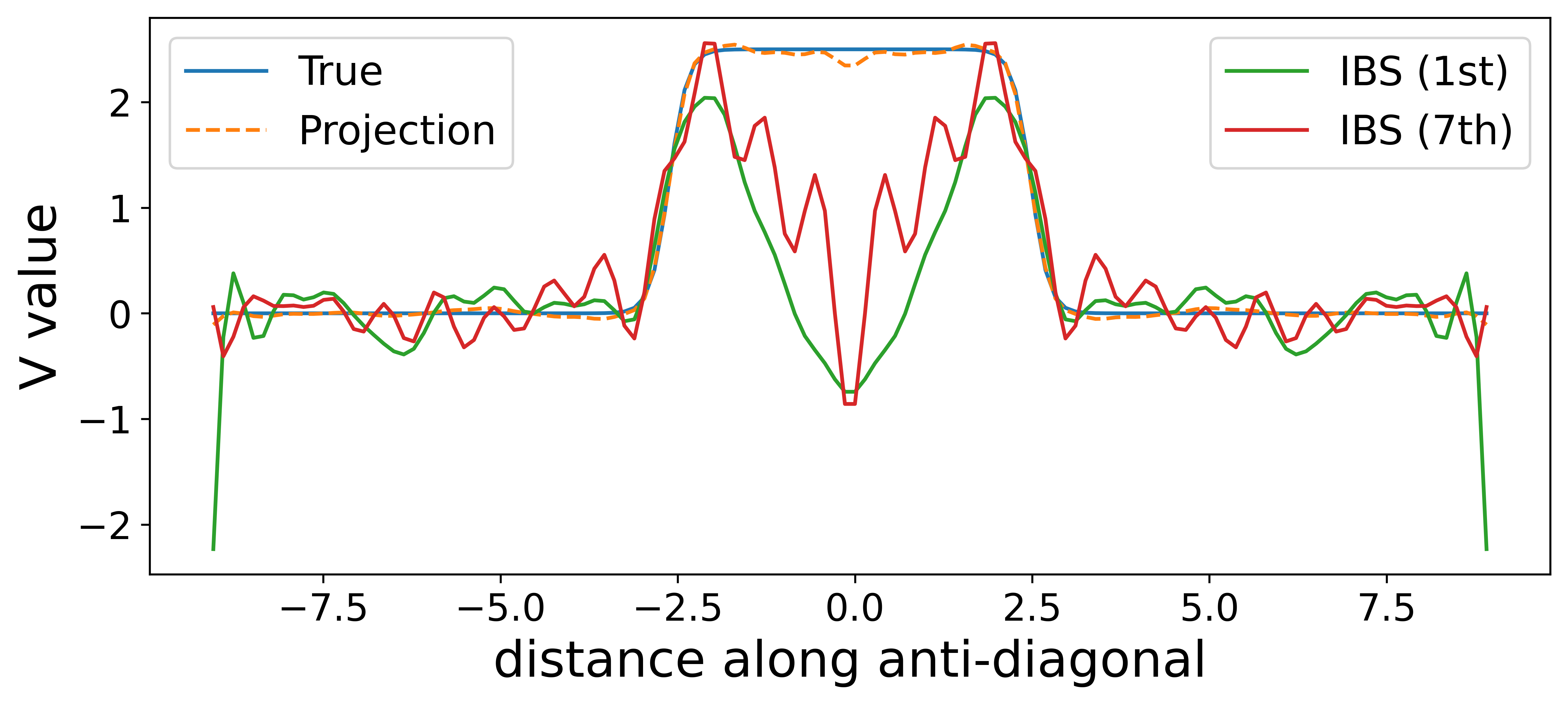}
\caption{Anti-diagonal cross section}
\end{subfigure}

\caption{Reconstructions of the high-contrast disk using the Fourier method.}
\label{fig:fourier-bad-circle}
\end{figure}

\begin{figure}[htbp]
\centering

\begin{subfigure}[b]{0.95\textwidth}
\includegraphics[width=\textwidth]{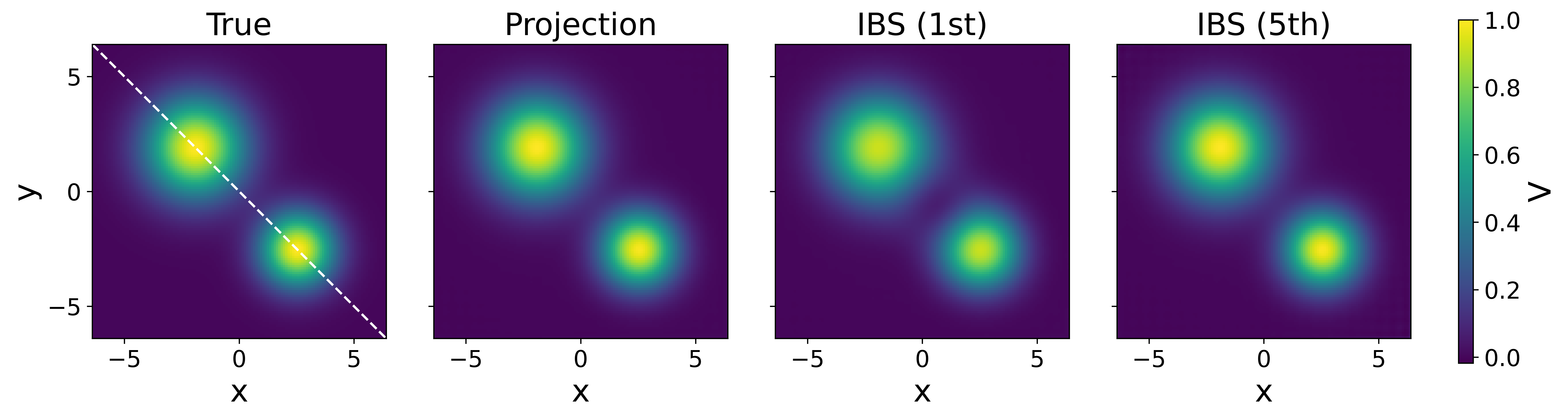}
\caption{Reconstruction of $V$ using phase data}
\end{subfigure}

\begin{subfigure}[b]{0.50\textwidth}
\includegraphics[width=\textwidth]{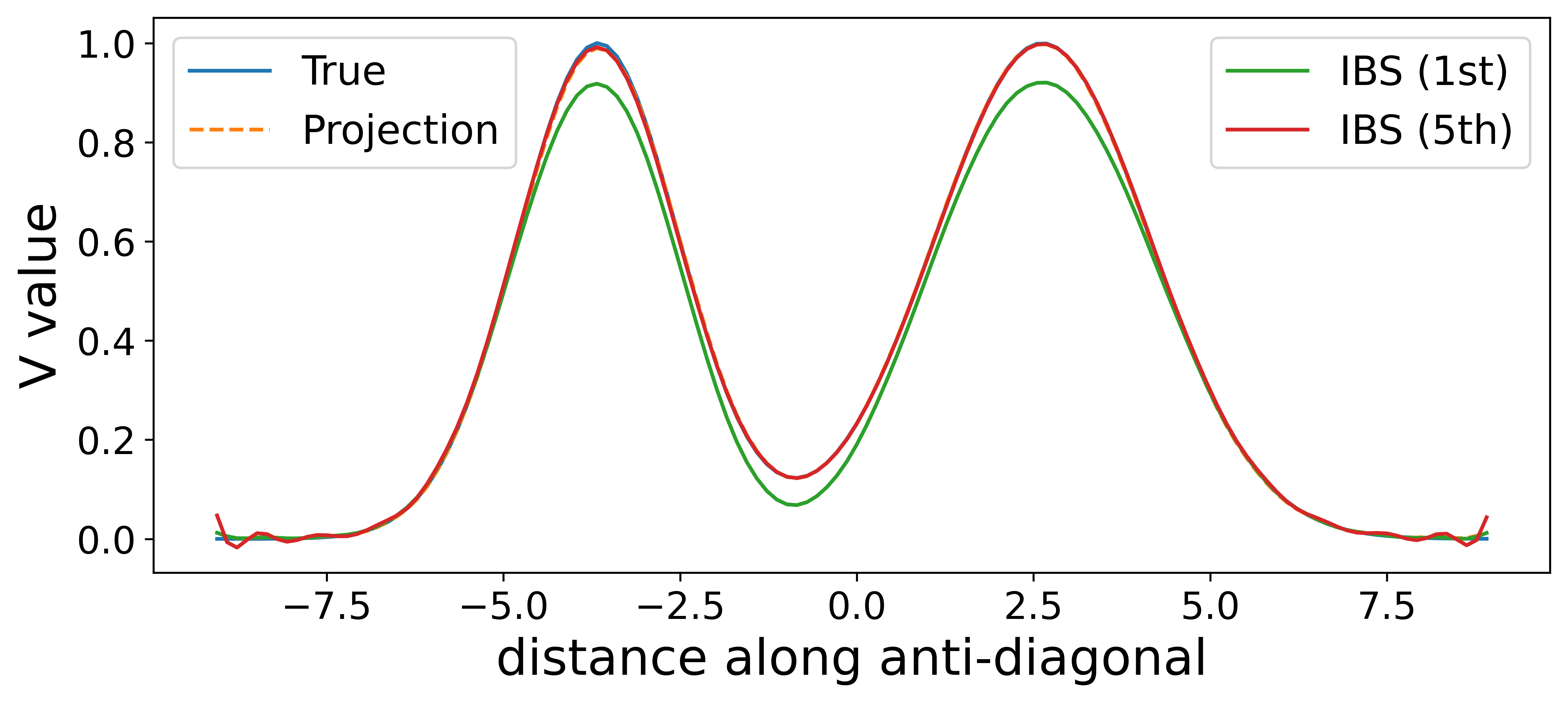}
\caption{Anti-diagonal cross section}
\end{subfigure}

\begin{subfigure}[b]{0.95\textwidth}
\includegraphics[width=\textwidth]{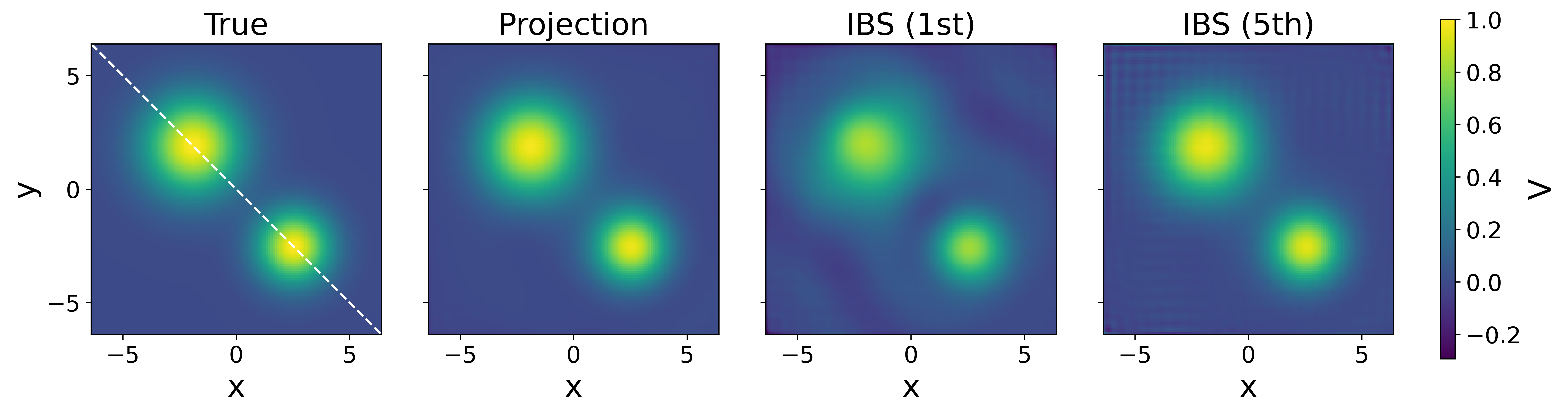}
\caption{Reconstruction of $V$ using phaseless data}
\end{subfigure}

\begin{subfigure}[b]{0.50\textwidth}
\includegraphics[width=\textwidth]{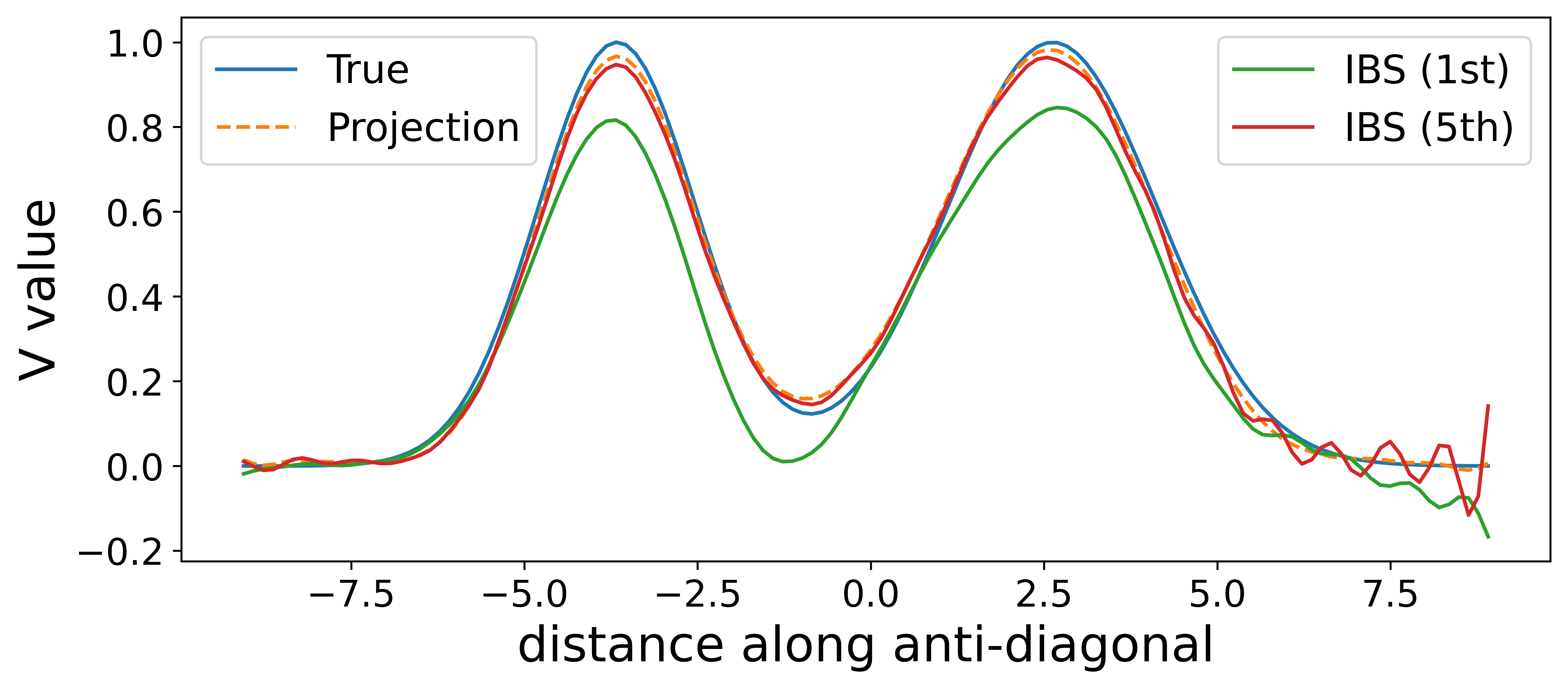}
\caption{Anti-diagonal cross section}
\end{subfigure}

\caption{Reconstructions of the low-contrast Gaussian mixture using the Fourier method.}
\label{fig:fourier-good-gaussian}
\end{figure}

\begin{figure}[htbp]
\centering

\begin{subfigure}[b]{0.95\textwidth}
\includegraphics[width=\textwidth]{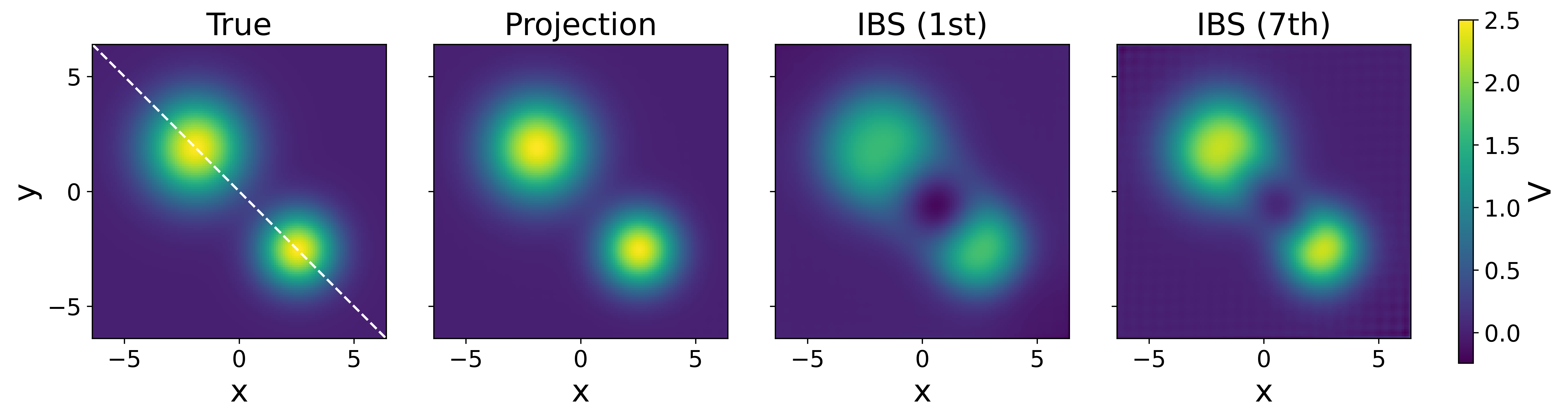}
\caption{Reconstruction of $V$ using phase data}
\end{subfigure}

\begin{subfigure}[b]{0.49\textwidth}
\includegraphics[width=\textwidth]{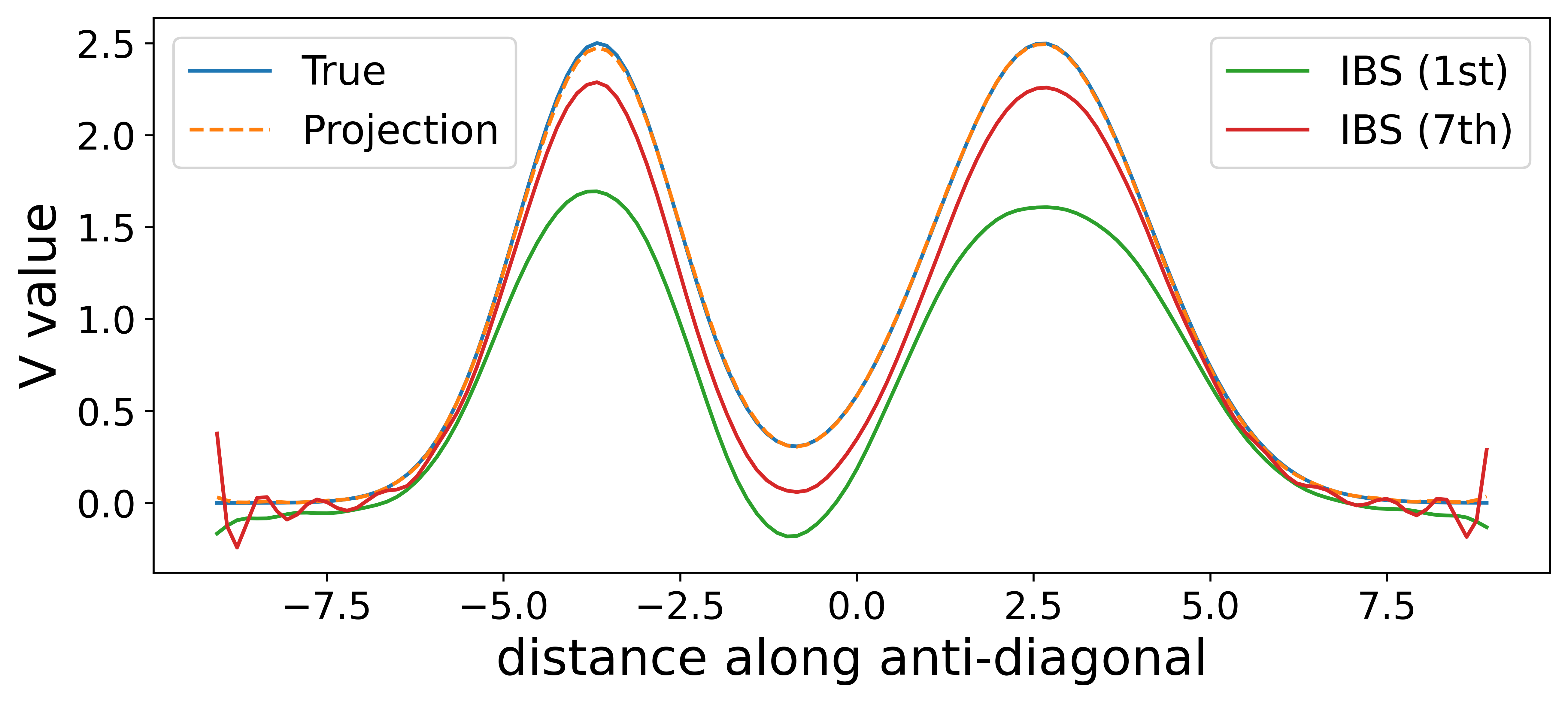}
\caption{Anti-diagonal cross section}
\end{subfigure}

\begin{subfigure}[b]{0.95\textwidth}
\includegraphics[width=\textwidth]{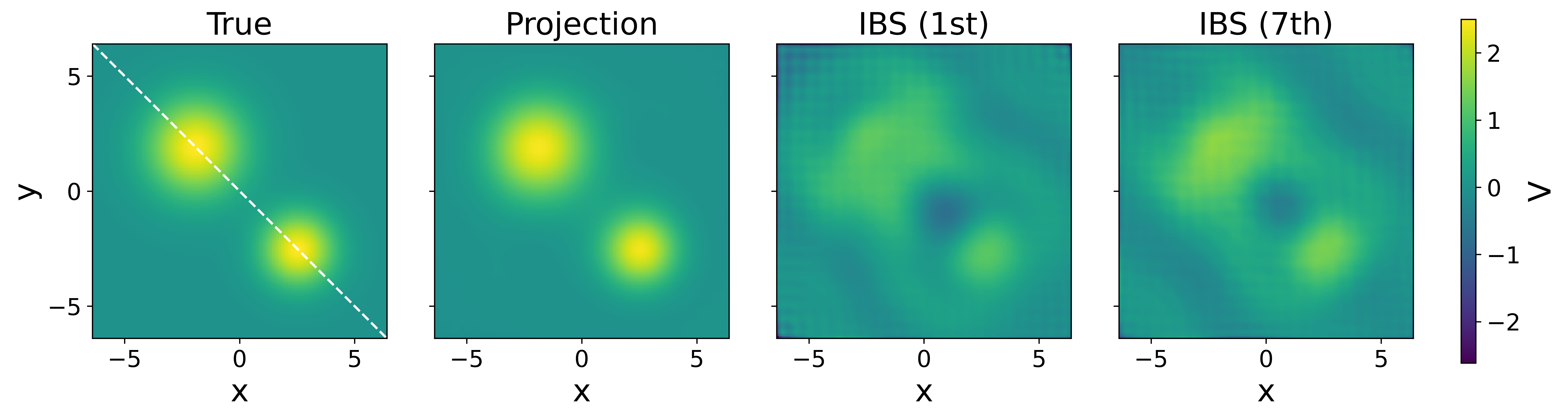}
\caption{Reconstruction of $V$ using phaseless data}
\end{subfigure}

\begin{subfigure}[b]{0.49\textwidth}
\includegraphics[width=\textwidth]{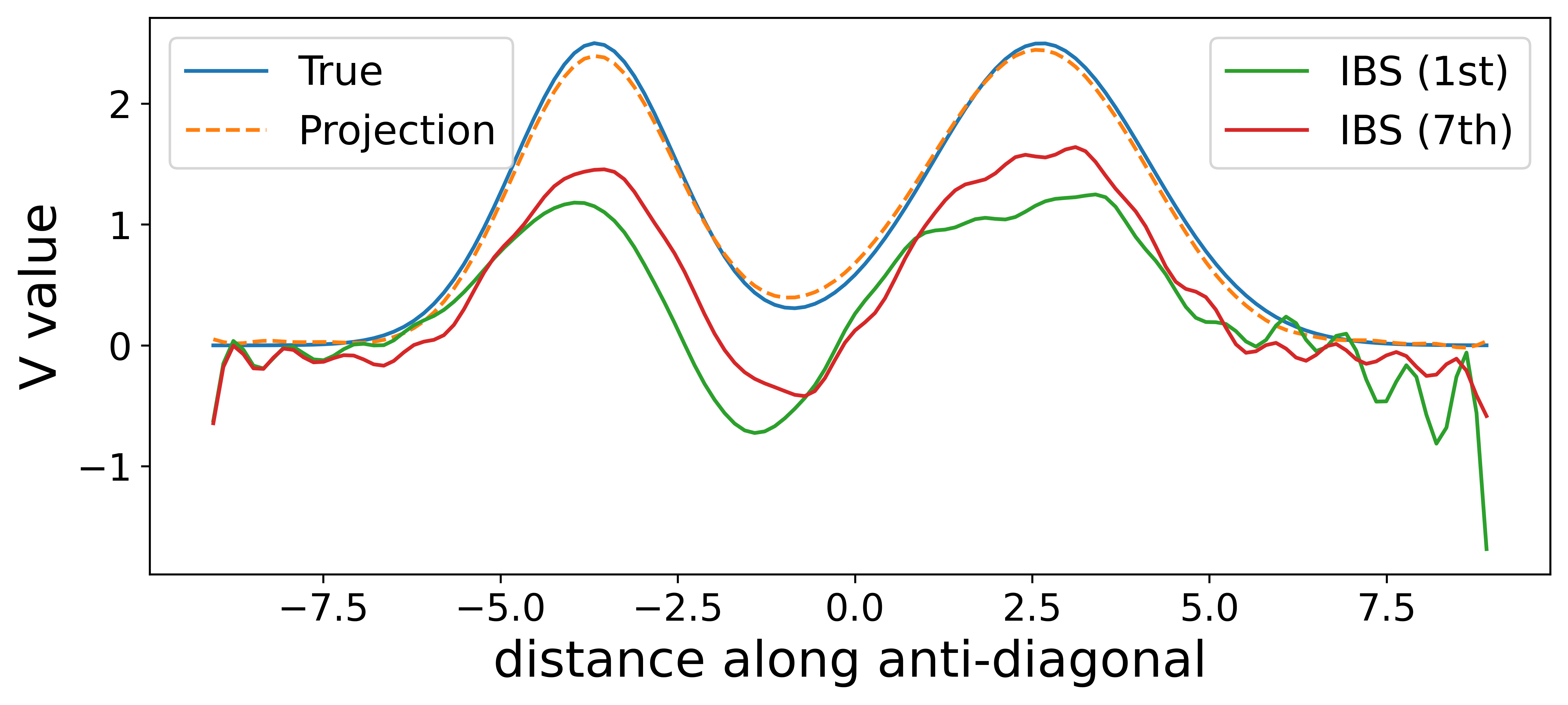}
\caption{Anti-diagonal cross section}
\end{subfigure}

\caption{Reconstructions of the high-contrast Gaussian mixture using the Fourier method.}
\label{fig:fourier-bad-gaussian}
\end{figure}

In Figures \ref{fig:fourier-good-circle} and \ref{fig:fourier-bad-circle}, we present the numerical results for low- and high-contrast disk potentials, with amplitudes $A = 1.0$ and $A = 2.5$, respectively. 
In Figures \ref{fig:fourier-good-gaussian} and \ref{fig:fourier-bad-gaussian}, we present the numerical results for low- and high-contrast Gaussian mixture potentials, with amplitudes $A = 1.0$ and $A = 2.5$, respectively. 

As expected from the theoretical analysis, when the contrast is low, the method works well, and the reconstructions from phase data are slightly better than those from phaseless data. 
When the contrast is high, for both examples, the IBS iteration for phase data still works well, while that for phaseless data fails. This is consistent with our analysis that the IBS for phase data has a larger radius of convergence.

\begin{table}[htbp]
\centering
\caption{Relative reconstruction errors corresponding to
Fig.~\ref{fig:fourier-good-circle}.}
\label{tab:fourier-good-circle}
\small
\renewcommand{\arraystretch}{1.15}
\setlength{\tabcolsep}{4pt}
\begin{tabular}{@{}lcccccc@{}}
\toprule
 & Projection & IBS1 & IBS2 & IBS3 & IBS4 & IBS5 \\
\midrule
Phase data
& 0.0121 & 0.0614 & 0.0411 & 0.0133 & 0.0131 & 0.0123 \\
Phaseless total-field data
& 0.0345 & 0.2215 & 0.1164 & 0.0694 & 0.0630 & 0.0594 \\
Phaseless scattered-field data
& 0.0121 & 0.0652 & 0.0429 & 0.0172 & 0.0167 & 0.0160 \\
\bottomrule
\end{tabular}
\end{table}

\begin{table}[htbp]
\centering
\caption{Relative reconstruction errors corresponding to
Fig.~\ref{fig:fourier-bad-circle}.}
\label{tab:fourier-bad-circle}
\small
\renewcommand{\arraystretch}{1.15}
\setlength{\tabcolsep}{4pt}
\begin{tabular}{@{}lcccccc@{}}
\toprule
 & Projection & IBS1 & IBS3 & IBS5 & IBS7 & IBS9 \\
\midrule
Phase data
& 0.0121 & 0.2609 & 0.1235 & 0.0765 & 0.0529 & 0.0391 \\
Phaseless total-field data
& 0.0345 & 0.6049 & 0.4719 & 0.4297 & 0.4398 & -- \\
Phaseless scattered-field data
& 0.0121 & 0.2571 & 0.1130 & 0.0655 & 0.0469 & 0.0416 \\
\bottomrule
\end{tabular}
\end{table}

\begin{table}[htbp]
\centering
\caption{Relative reconstruction errors corresponding to
Fig.~\ref{fig:fourier-good-gaussian}.}
\label{tab:fourier-good-gaussian}
\small
\renewcommand{\arraystretch}{1.15}
\setlength{\tabcolsep}{4pt}
\begin{tabular}{@{}lcccccc@{}}
\toprule
 & Projection & IBS1 & IBS2 & IBS3 & IBS4 & IBS5 \\
\midrule
Phase data
& 0.0082 & 0.0582 & 0.0412 & 0.0120 & 0.0103 & 0.0099 \\
Phaseless total-field data
& 0.0598 & 0.2270 & 0.1248 & 0.0941 & 0.0887 & 0.0865 \\
Phaseless scattered-field data
& 0.0082 & 0.0660 & 0.0473 & 0.0242 & 0.0222 & 0.0212 \\
\bottomrule
\end{tabular}
\end{table}

\begin{table}[htbp]
\centering
\caption{Relative reconstruction errors corresponding to
Fig.~\ref{fig:fourier-bad-gaussian}.}
\label{tab:fourier-bad-gaussian}
\small
\renewcommand{\arraystretch}{1.15}
\setlength{\tabcolsep}{4pt}
\begin{tabular}{@{}lcccccc@{}}
\toprule
 & Projection & IBS1 & IBS2 & IBS3 & IBS5 & IBS7 \\
\midrule
Phase data
& 0.0082 & 0.2609 & 0.2363 & 0.1478 & 0.1055 & 0.0833 \\
Phaseless total-field data
& 0.0607 & 0.5945 & 0.5225 & 0.4941 & 0.4614 & 0.4441 \\
Phaseless scattered-field data
& 0.0082 & 0.2640 & 0.2378 & 0.1571 & 0.1186 & 0.0990 \\
\bottomrule
\end{tabular}
\end{table}
\subsection{Polarization method}

In this subsection, we employ the same far-field setting as in Section \ref{sec:fourier}. 
For the polarization method, the same regularization parameter $\lambda = 10$ is used for both phase and phaseless data.

\begin{figure}[htbp]
\centering

\begin{subfigure}[b]{0.95\textwidth}
\includegraphics[width=\textwidth]{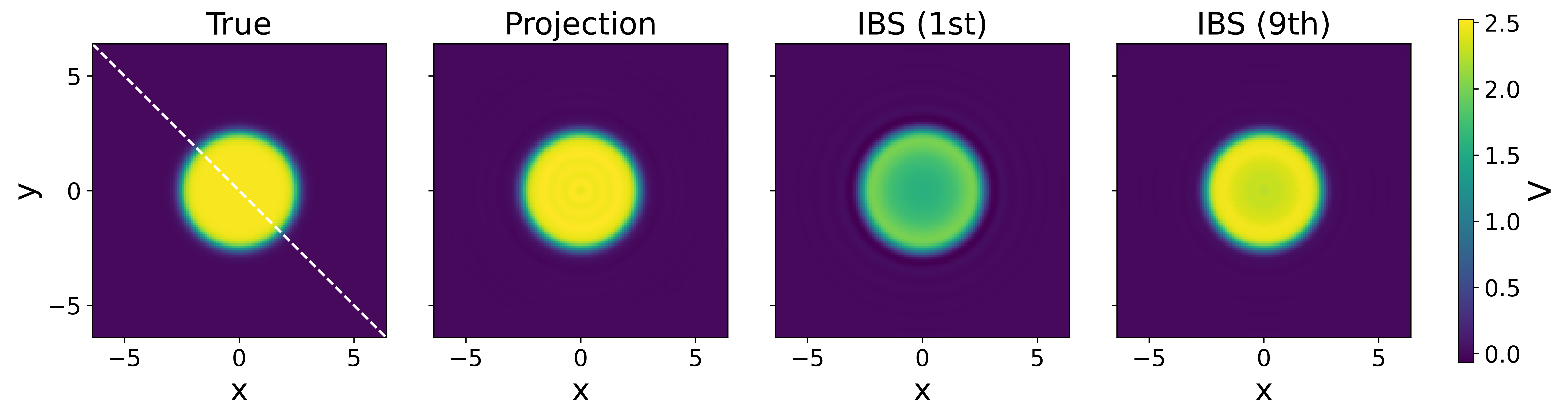}
\caption{Reconstruction of $V$ using phase data}
\label{fig:polar-good-circle-phase-4panel}
\end{subfigure}

\begin{subfigure}[b]{0.50\textwidth}
\includegraphics[width=\textwidth]{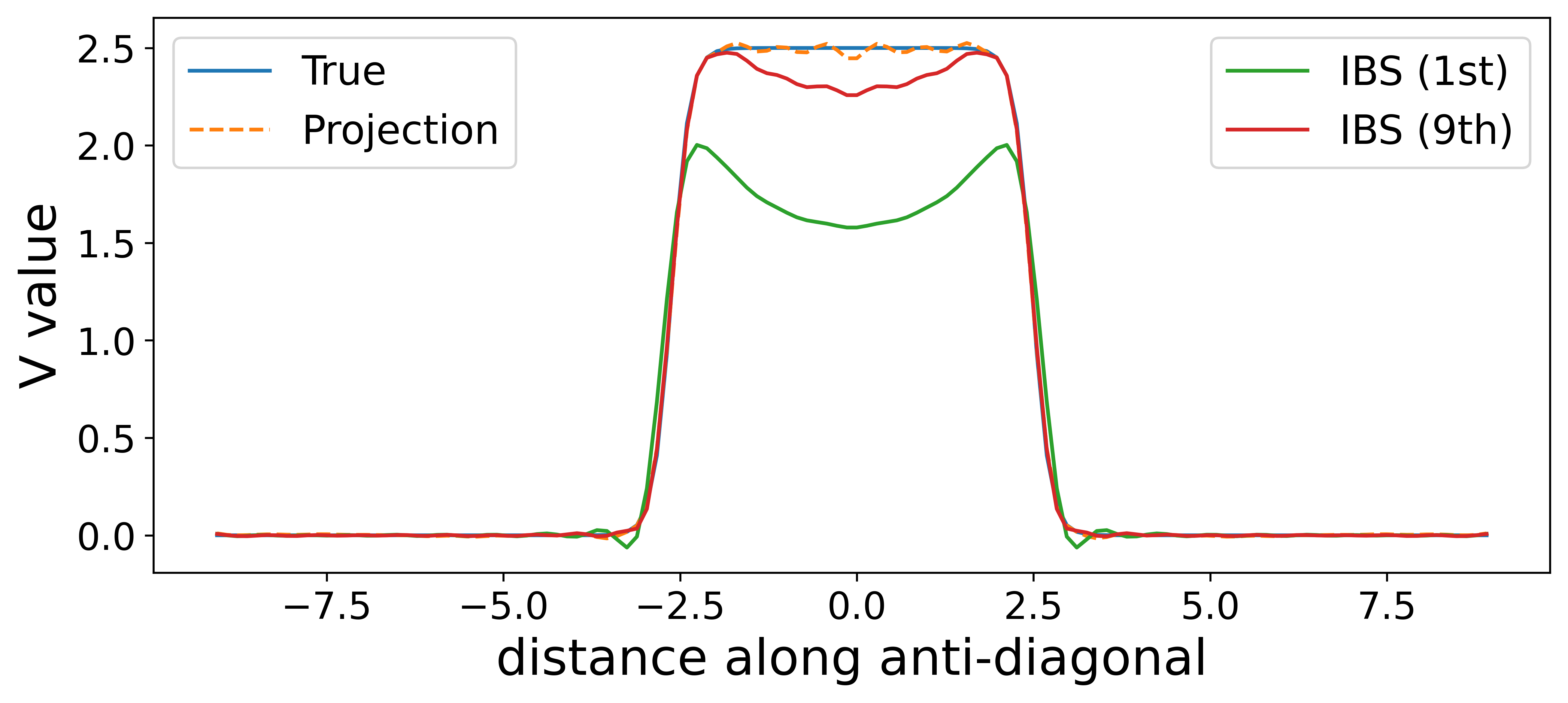}
\caption{Anti-diagonal cross section}
\label{fig:polar-good-circle-phase-slice}
\end{subfigure}

\begin{subfigure}[b]{0.95\textwidth}
\includegraphics[width=\textwidth]{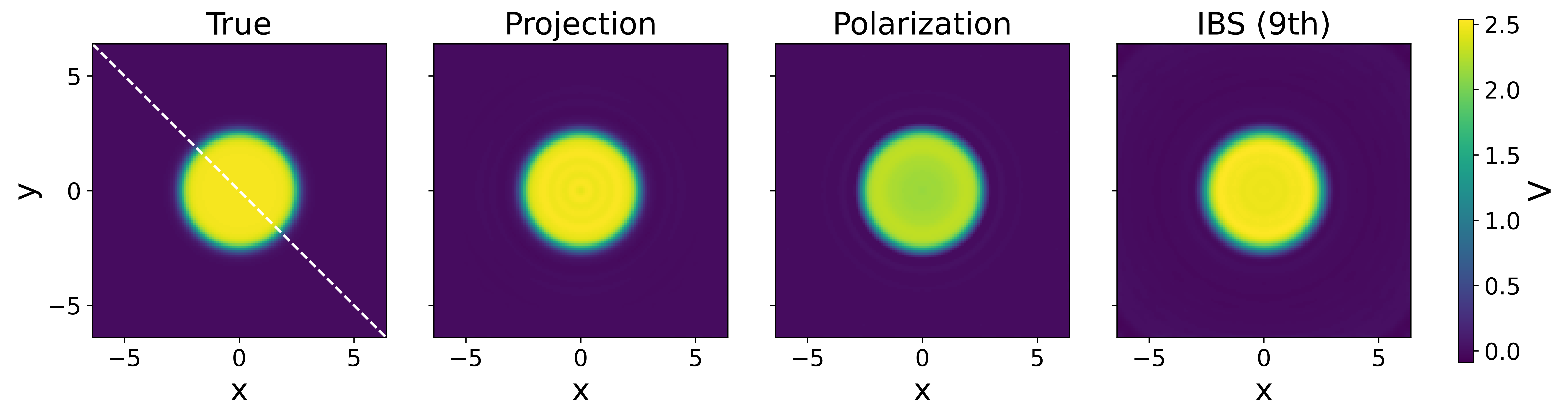}
\caption{Reconstruction of $V$ using phaseless data}
\label{fig:polar-good-circle-phaseless-4panel}
\end{subfigure}

\begin{subfigure}[b]{0.50\textwidth}
\includegraphics[width=\textwidth]{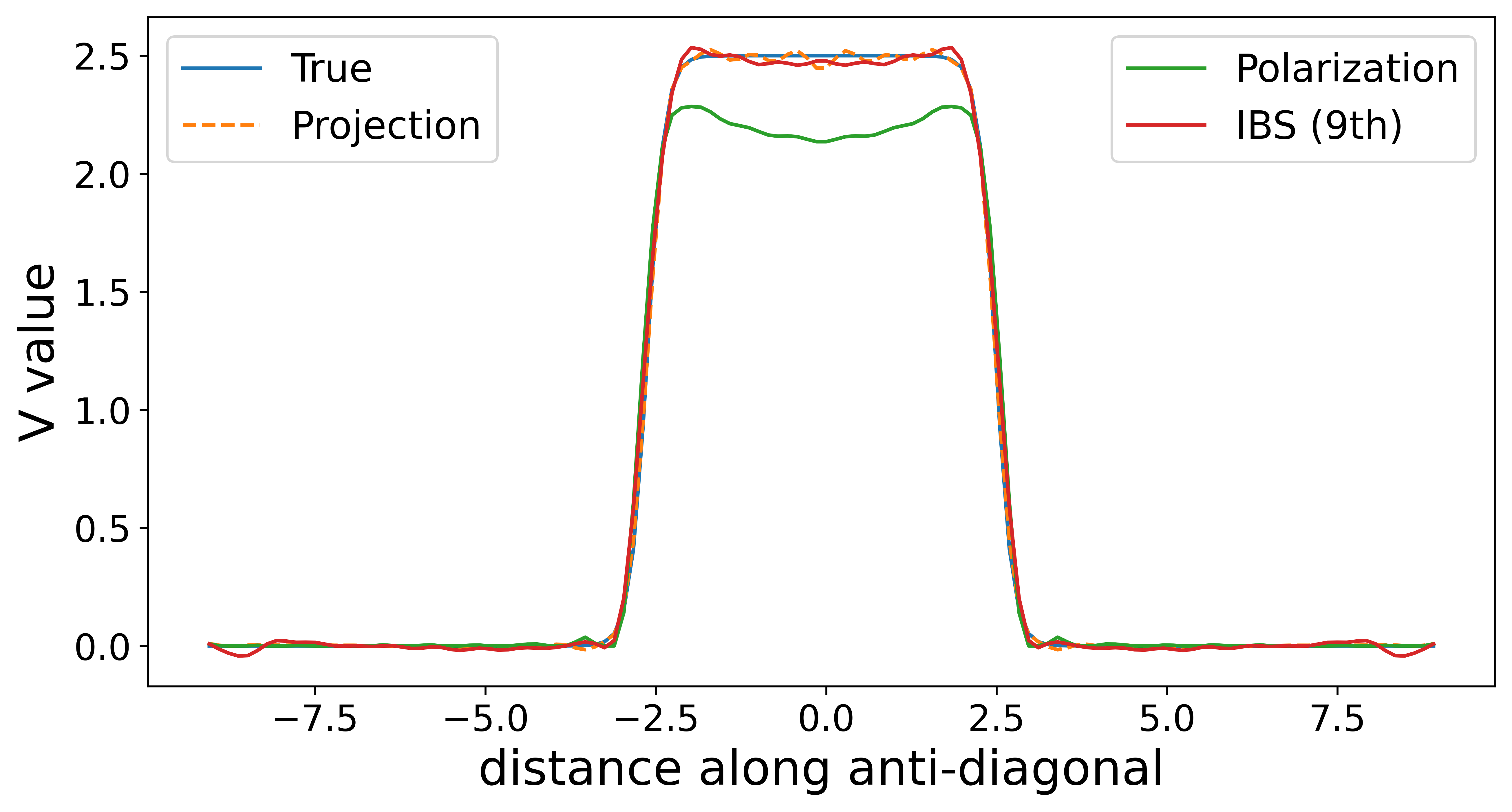}
\caption{Anti-diagonal cross section}
\label{fig:polar-good-circle-phaseless-slice}
\end{subfigure}

\caption{Reconstructions of the low-contrast disk using the polarization method.}
\label{fig:polar-good-circle}
\end{figure}

\begin{figure}[htbp]
\centering

\begin{subfigure}[b]{0.95\textwidth}
\includegraphics[width=\textwidth]{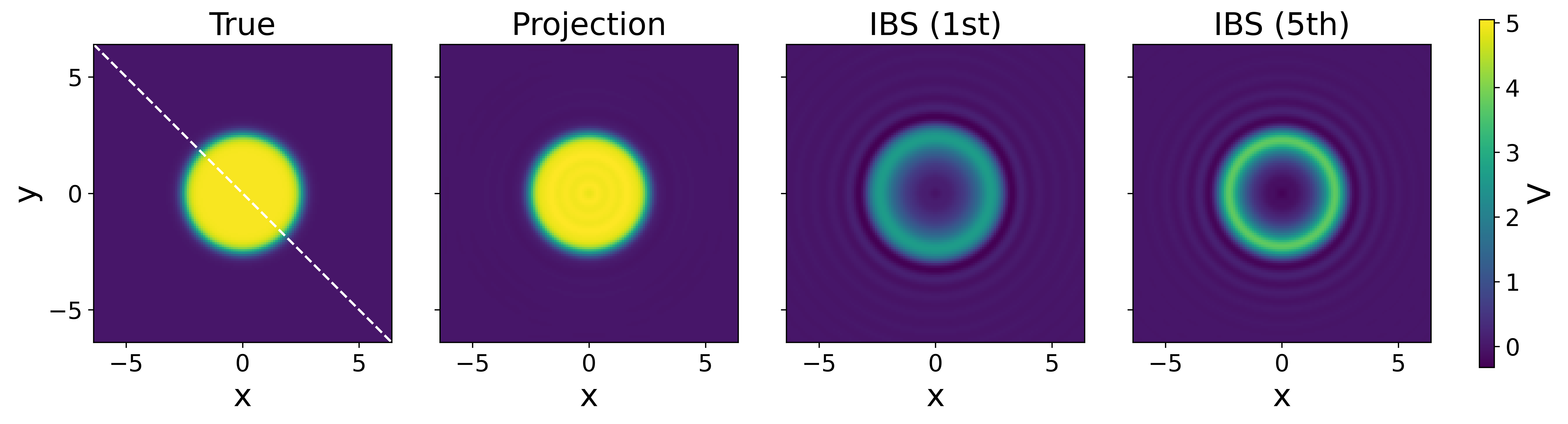}
\caption{Reconstruction of $V$ using phase data}
\label{fig:polar-bad-circle-phase-4panel}
\end{subfigure}

\begin{subfigure}[b]{0.50\textwidth}
\includegraphics[width=\textwidth]{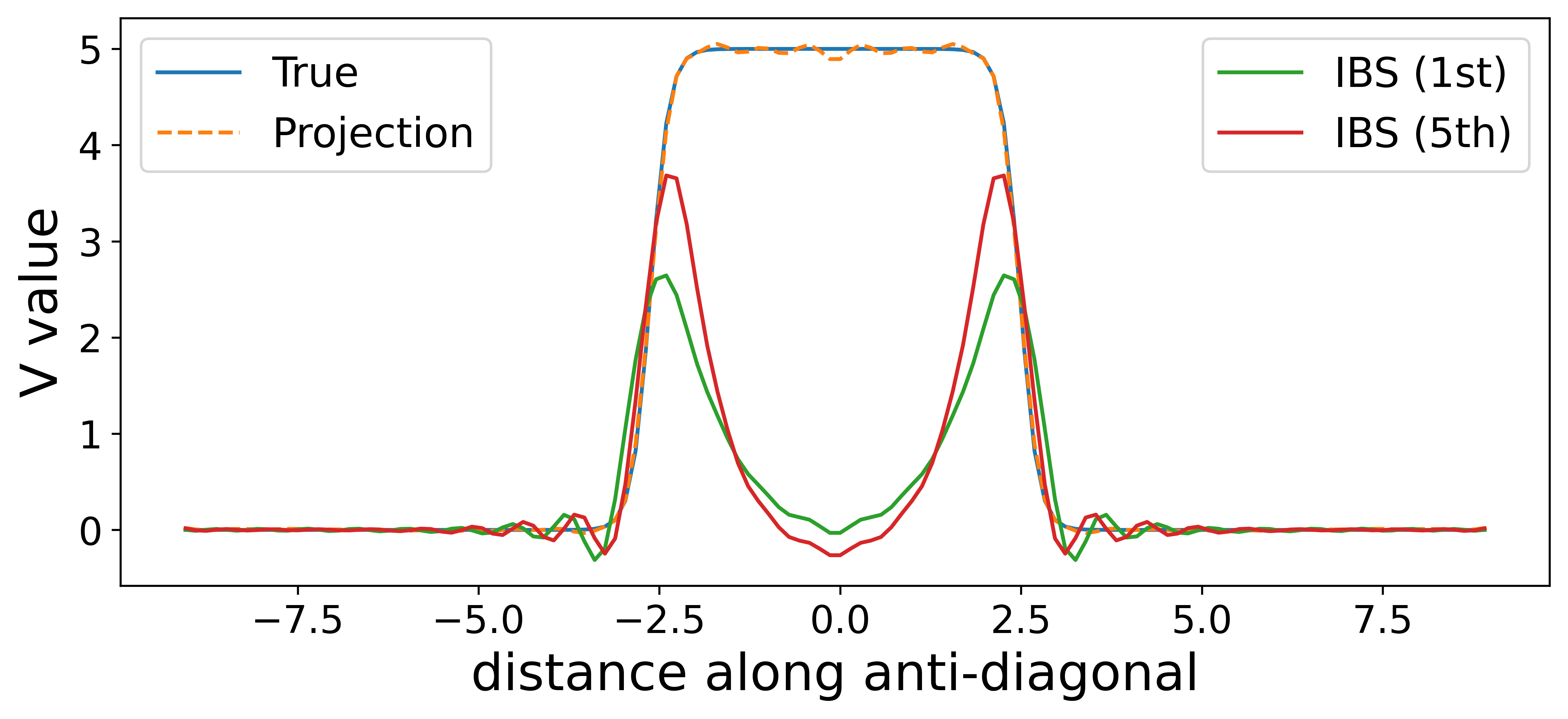}
\caption{Anti-diagonal cross section}
\label{fig:polar-bad-circle-phase-slice}
\end{subfigure}

\begin{subfigure}[b]{0.95\textwidth}
\includegraphics[width=\textwidth]{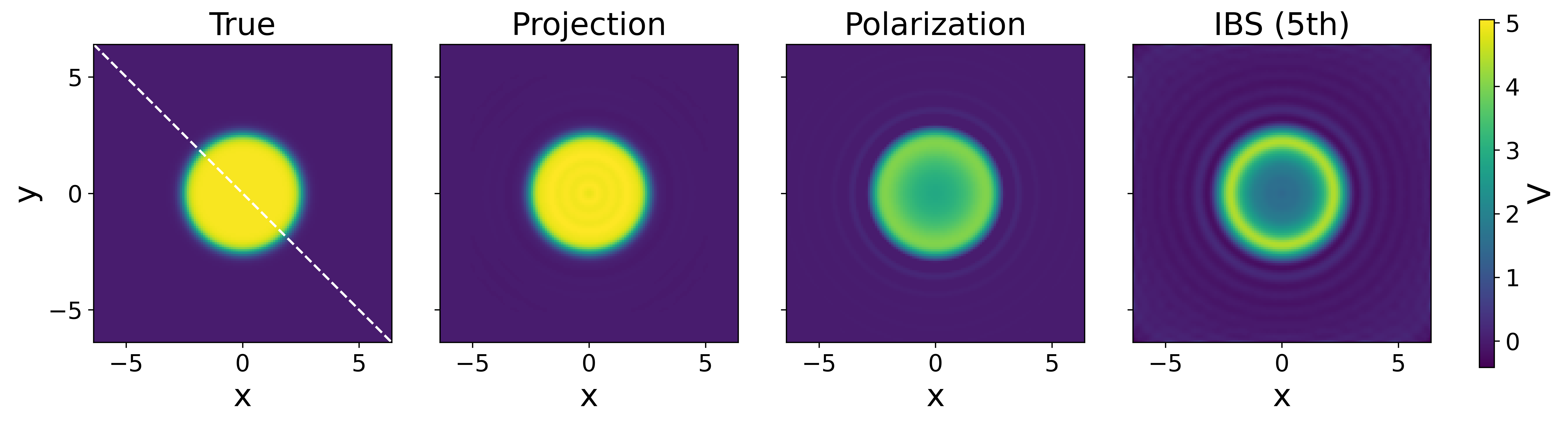}
\caption{Reconstruction of $V$ using phaseless data}
\label{fig:polar-bad-circle-phaseless-4panel}
\end{subfigure}

\begin{subfigure}[b]{0.50\textwidth}
\includegraphics[width=\textwidth]{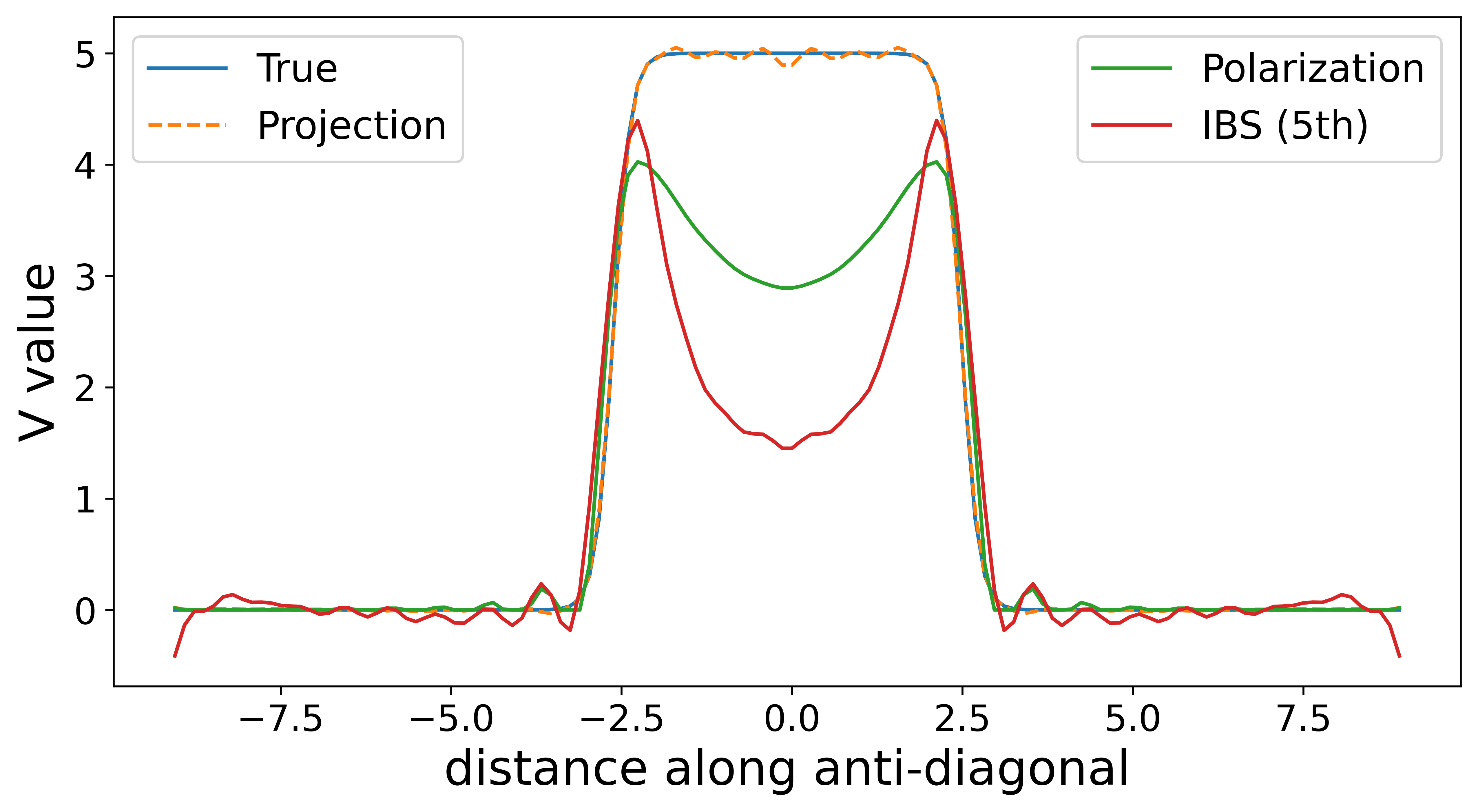}
\caption{Anti-diagonal cross section}
\label{fig:polar-bad-circle-phaseless-slice}
\end{subfigure}

\caption{Reconstructions of the high-contrast disk using the polarization method.}
\label{fig:polar-bad-circle}
\end{figure}

\begin{figure}[htbp]
\centering

\begin{subfigure}[b]{0.95\textwidth}
\includegraphics[width=\textwidth]{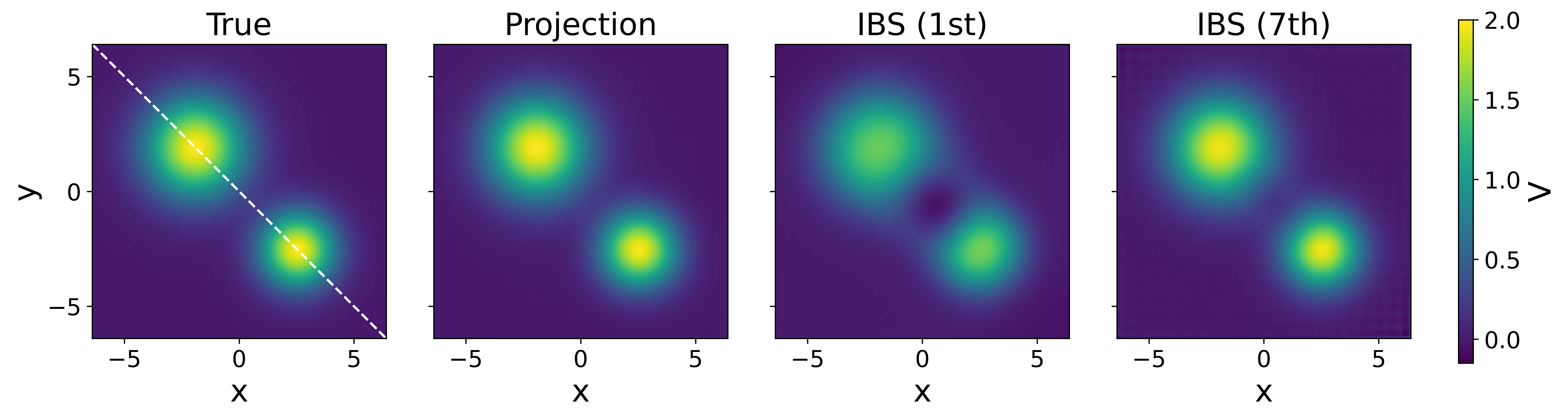}
\caption{Reconstruction of $V$ using phase data}
\label{fig:polar-good-gaussian-phase-4panel}
\end{subfigure}

\begin{subfigure}[b]{0.49\textwidth}
\includegraphics[width=\textwidth]{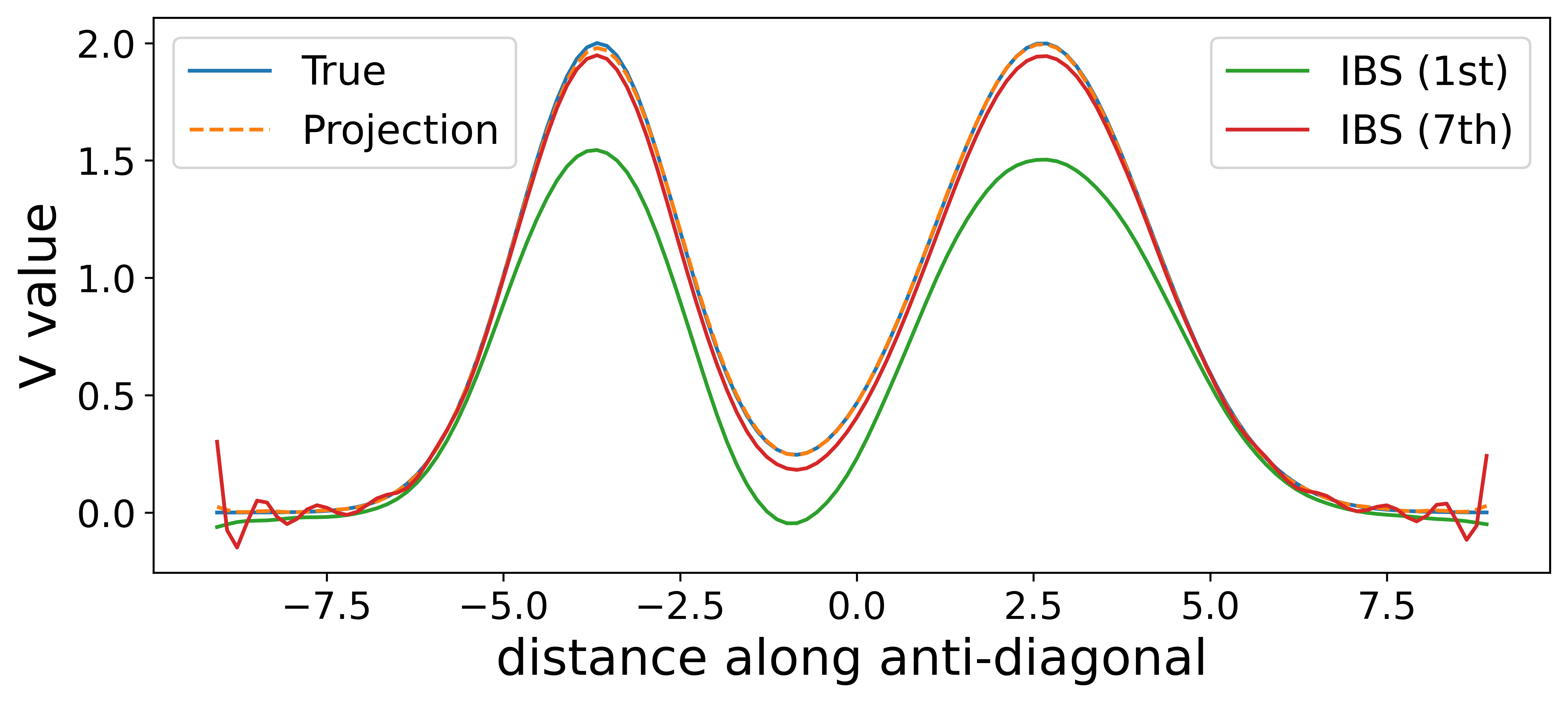}
\caption{Anti-diagonal cross section}
\label{fig:polar-good-gaussian-phase-slice}
\end{subfigure}

\begin{subfigure}[b]{0.95\textwidth}
\includegraphics[width=\textwidth]{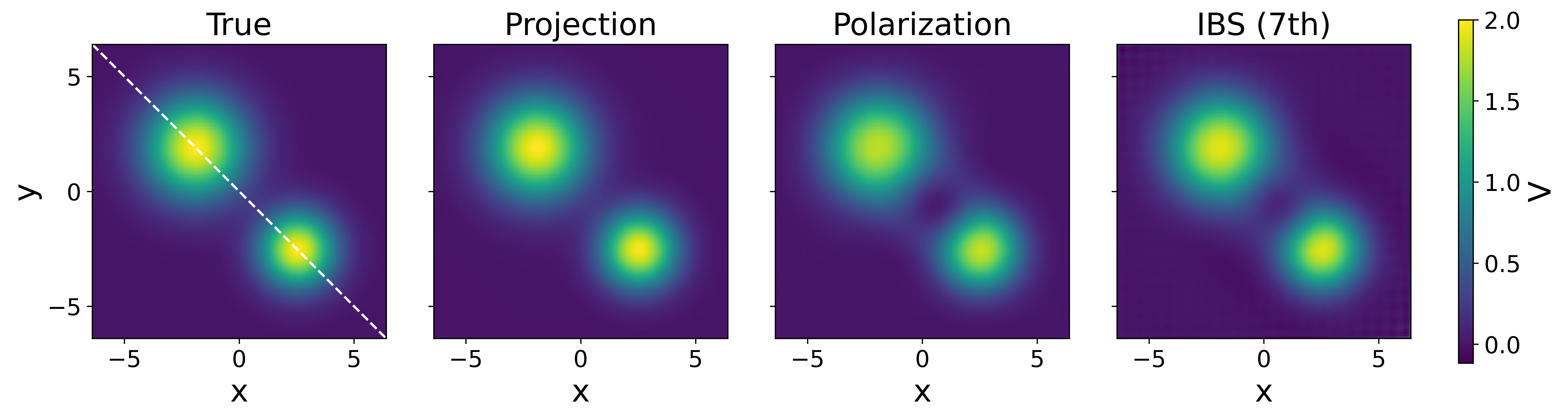}
\caption{Reconstruction of $V$ using phaseless data}
\label{fig:polar-good-gaussian-phaseless-4panel}
\end{subfigure}

\begin{subfigure}[b]{0.49\textwidth}
\includegraphics[width=\textwidth]{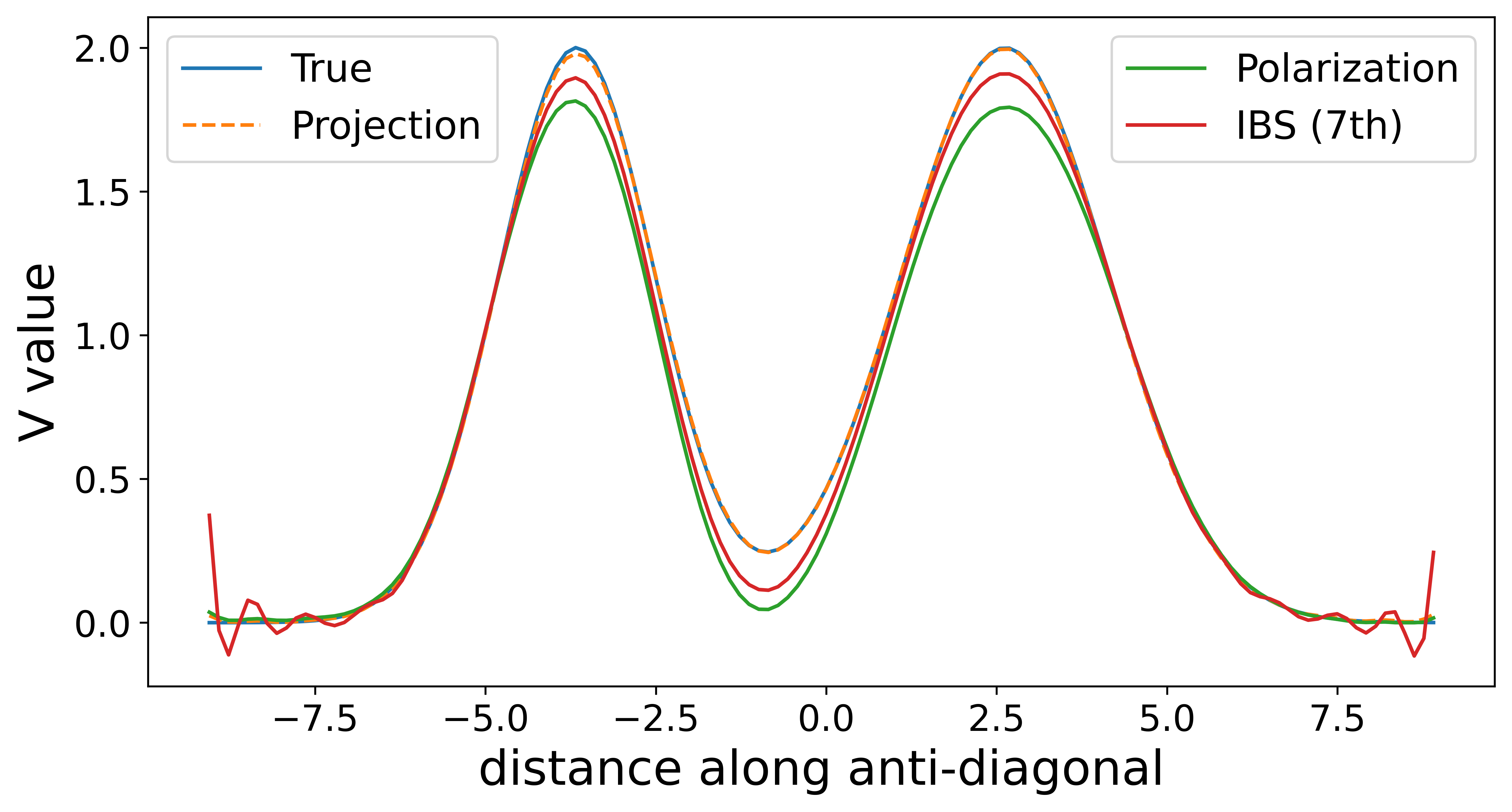}
\caption{Anti-diagonal cross section}
\label{fig:polar-good-gaussian-phaseless-slice}
\end{subfigure}

\caption{Reconstructions of the low-contrast Gaussian mixture using the polarization method.}
\label{fig:polar-good-gaussian}
\end{figure}

\begin{figure}[htbp]
\centering

\begin{subfigure}[b]{0.95\textwidth}
\includegraphics[width=\textwidth]{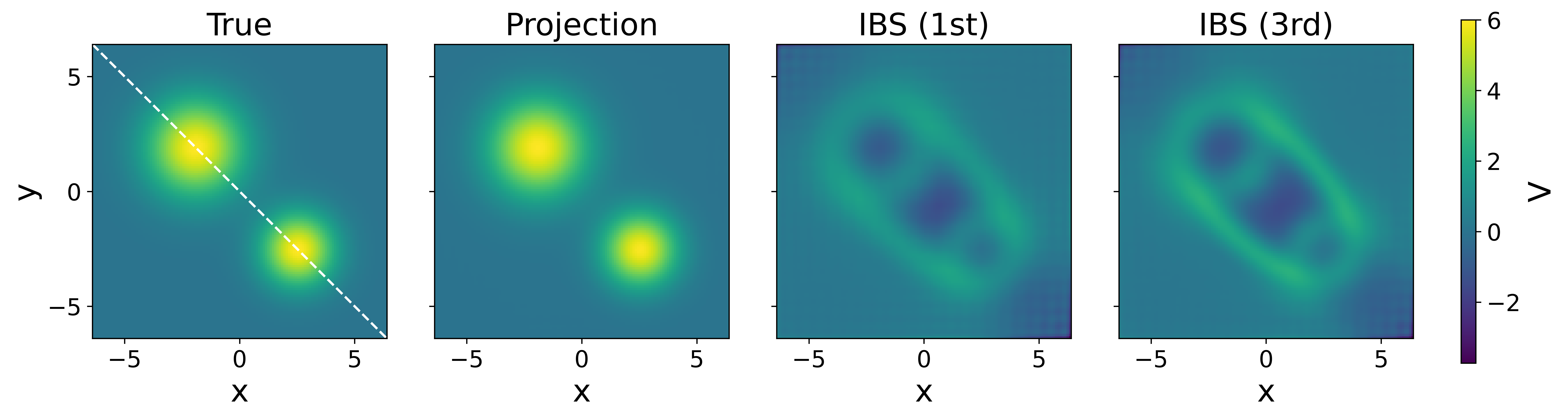}
\caption{Reconstruction of $V$ using phase data}
\label{fig:polar-bad-gaussian-phase-4panel}
\end{subfigure}

\begin{subfigure}[b]{0.50\textwidth}
\includegraphics[width=\textwidth]{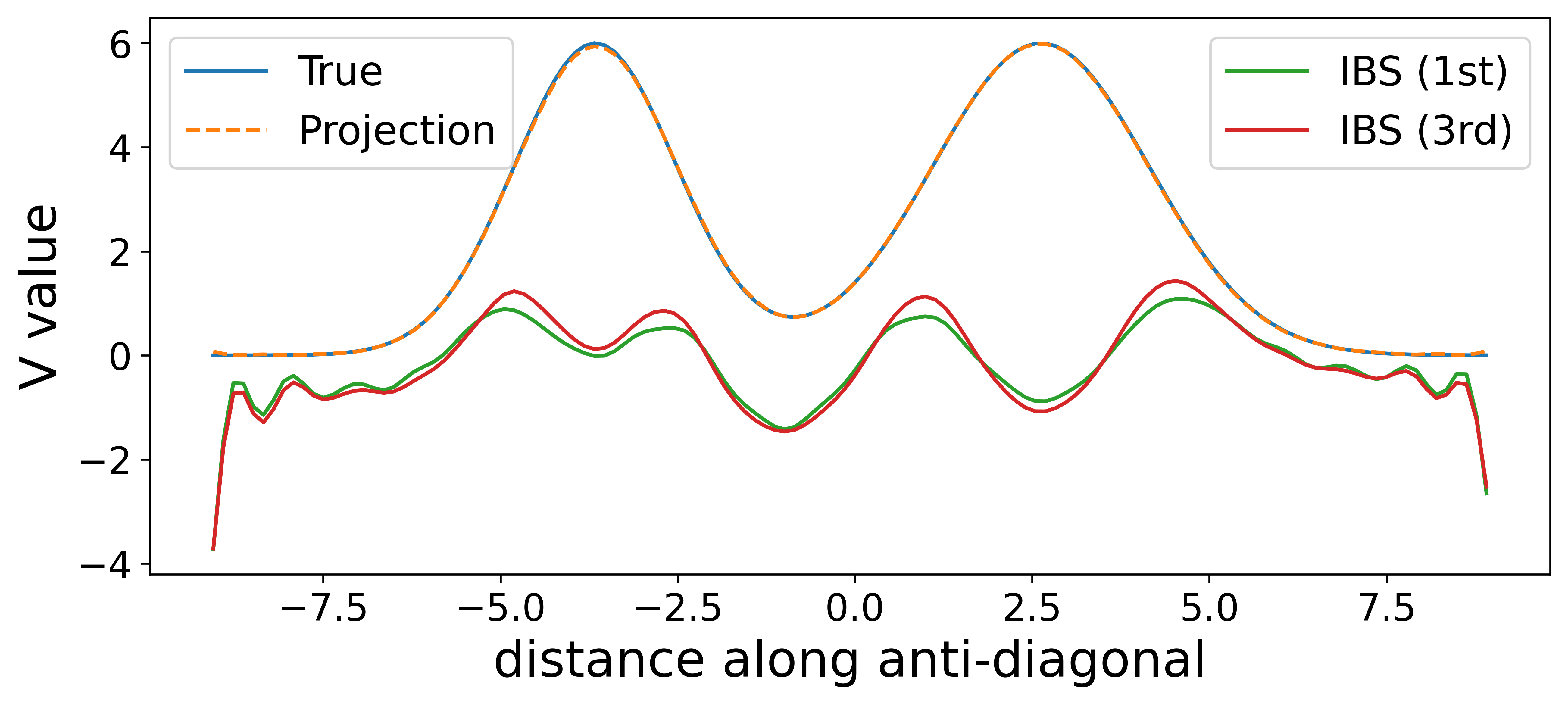}
\caption{Anti-diagonal cross section}
\label{fig:polar-bad-gaussian-phase-slice}
\end{subfigure}

\begin{subfigure}[b]{0.95\textwidth}
\includegraphics[width=\textwidth]{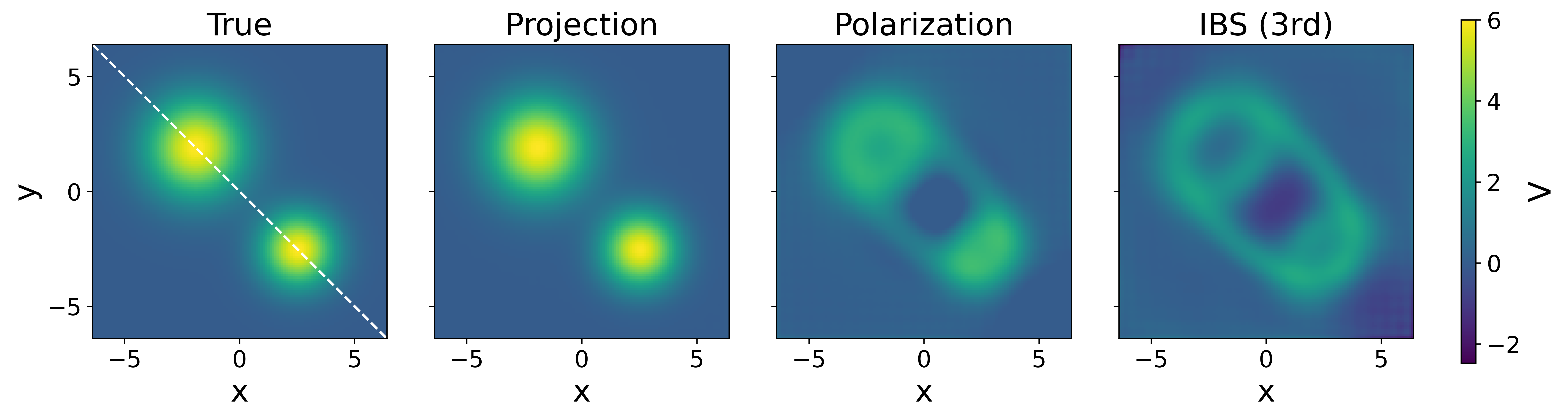}
\caption{Reconstruction of $V$ using phaseless data}
\label{fig:polar-bad-gaussian-phaseless-4panel}
\end{subfigure}

\begin{subfigure}[b]{0.50\textwidth}
\includegraphics[width=\textwidth]{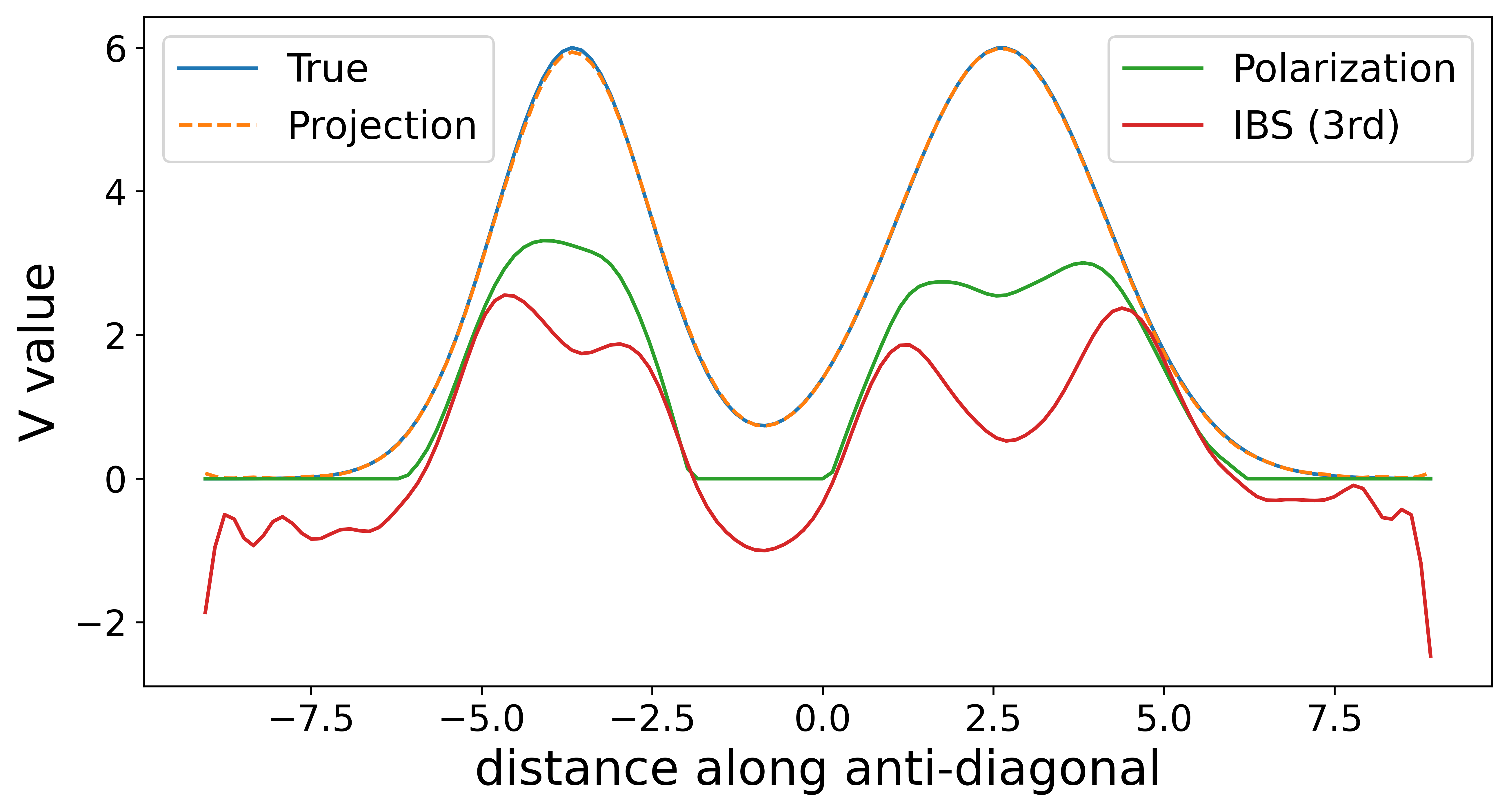}
\caption{Anti-diagonal cross section}
\label{fig:polar-bad-gaussian-phaseless-slice}
\end{subfigure}

\caption{Reconstructions of the high-contrast Gaussian mixture using the polarization method.}
\label{fig:polar-bad-gaussian}
\end{figure}

In Figures \ref{fig:polar-good-circle} and \ref{fig:polar-bad-circle}, we present the numerical results for low- and high-contrast disk potentials, with amplitudes $A = 2.5$ and $A = 5.0$, respectively. 
In Figures \ref{fig:polar-good-gaussian} and \ref{fig:polar-bad-gaussian}, we present the numerical results for low- and high-contrast Gaussian mixture potentials, with amplitudes $A = 2.0$ and $A = 6.0$, respectively. 

As expected from the theoretical analysis, when the contrast is low, the method works well, and the reconstructions from phase data are slightly better than those from phaseless data. 
When the contrast is high, the IBS iteration fails to converge in both cases. 

We also compare reconstructions from phaseless data of the total field and from phaseless data of the scattered field. 
For the same scattering potential, the corresponding reconstruction errors are reported in the tables below Figures \ref{fig:fourier-good-circle}, \ref{fig:fourier-bad-circle}, \ref{fig:fourier-good-gaussian}, and \ref{fig:fourier-bad-gaussian}. 
As expected, the reconstruction from phase data is the most accurate. 
Moreover, the reconstruction from phaseless scattered-field data is better than from phaseless total-field data. The reason is that, in the reconstruction from total-field data, some Fourier samples must be discarded, as stated in Algorithm 2(4), even though the phase is approximated in the polarization method, as described in Algorithm 3(3).

\begin{table}[htbp]
\centering
\caption{Relative reconstruction errors corresponding to
Fig.~\ref{fig:polar-good-circle}.}
\label{tab:polar-good-circle}
\small
\renewcommand{\arraystretch}{1.15}
\setlength{\tabcolsep}{4pt}
\begin{tabular}{@{}lccccccc@{}}
\toprule
 & Projection & Polarization & IBS1 & IBS3 & IBS5 & IBS7 & IBS9 \\
\midrule
Phase data
& 0.0121 & -- & 0.2609 & 0.1235 & 0.0765 & 0.0529 & 0.0391 \\
Phaseless data
& 0.0121 & 0.1092 & 0.2571 & 0.1130 & 0.0655 & 0.0469 & 0.0416 \\
\bottomrule
\end{tabular}
\end{table}

\begin{table}[htbp]
\centering
\caption{Relative reconstruction errors corresponding to
Fig.~\ref{fig:polar-good-gaussian}.}
\label{tab:polar-good-gaussian}
\small
\renewcommand{\arraystretch}{1.15}
\setlength{\tabcolsep}{4pt}
\begin{tabular}{@{}lccccccc@{}}
\toprule
 & Projection & Polarization & IBS1 & IBS2 & IBS3 & IBS5 & IBS7 \\
\midrule
Phase data
& 0.0082 & -- & 0.1805 & 0.1563 & 0.0754 & 0.0439 & 0.0318 \\
Phaseless data
& 0.0082 & 0.0785 & 0.1899 & 0.1642 & 0.0937 & 0.0665 & 0.0559 \\
\bottomrule
\end{tabular}
\end{table}

\FloatBarrier
\section{Conclusion}\label{sec:con}
We have investigated the inverse scattering problem with phaseless data using the IBS method. 
We present three numerical methods for different types of data: the phaseless total field, the far-field phaseless total field, and the far-field phaseless scattered field. 
We analyzed the convergence of the IBS in all cases. For the phaseless IBS with total-field intensity data, we also develop an efficient recursive implementation of the multilinear operators \(K_n\), reducing the cost of evaluating each \(K_n\) from \(\mathcal{O}(n^2)\) to \(\mathcal{O}(n)\) convolution operations.
Numerical experiments were conducted to validate the methods. 
We also compare the proposed methods with each other, as well as with the corresponding cases when phase data are available.

We have found that all the methods perform well for low-contrast potentials and fail for high-contrast potentials, which is consistent with the analysis of the IBS. 
We also find that, in general, reconstructions from phase data are more accurate than those from phaseless data and the corresponding IBS exhibit a larger radius of convergence. 
Finally, the method for the far-field phaseless scattered field is superior to that for the far-field phaseless total field. This follows because the scattered-field setting contains more informative data, while the total-field approach requires discarding part of the Fourier information.

In future work, we plan to investigate other phaseless inverse scattering problems. These include the wave equations of elasticity and the nonlocal equations that arise in single-photon scattering. 
\bibliographystyle{siam}
\bibliography{ref}
\end{document}